\documentclass[aps,prc,twocolumn,superscriptaddress,floatfix,amsmath,amssymb,nofootinbib,longbibliography]{revtex4-2}
\usepackage[utf8]{inputenc}
\usepackage[T1]{fontenc}

\usepackage{xcolor}
\usepackage{graphicx}
\usepackage{microtype}
\usepackage{siunitx}
\usepackage{dcolumn}

\usepackage{braket}
\usepackage{booktabs}
\usepackage[colorlinks=true,citecolor=blue,linkcolor=blue,urlcolor=blue]{hyperref}

\usepackage{orcidlink}

\newcommand{\magic}{1.8/2.0~(EM)}

\newcommand{\MeV}{\text{MeV}}
\newcommand{\fminv}{\text{fm}^{-1}}
\newcommand{\fm}{\text{fm}}
\newcommand{\elem}[2]{\ensuremath{^{\text{#2}}\text{#1}}}

\newcommand{\nn}{\ensuremath{\text{NN}}}
\newcommand{\nnn}{\ensuremath{\text{3N}}}

\begin{document}

\title{Uncertainties with low-resolution nuclear forces}
\thanks{This manuscript has been authored in part by UT-Battelle, LLC, under contract DE-AC05-00OR22725 with the US Department of Energy (DOE). The US government retains and the publisher, by accepting the article for publication, acknowledges that the US government retains a nonexclusive, paid-up, irrevocable, worldwide license to publish or reproduce the published form of this manuscript, or allow others to do so, for US government purposes. DOE will provide public access to these results of federally sponsored research in accordance with the DOE Public Access Plan (\url{http://energy.gov/downloads/doe-public-access-plan}).}

\author{T.~Plies \orcidlink{0009-0002-7767-9659}}
\email{tom.plies@tu-darmstadt.de}
\affiliation{Technische Universit\"at Darmstadt, Department of Physics, 64289 Darmstadt, Germany}
\affiliation{ExtreMe Matter Institute EMMI, GSI Helmholtzzentrum f\"ur Schwerionenforschung GmbH, 64291 Darmstadt, Germany}

\author{M.~Heinz \orcidlink{0000-0002-6363-0056}}
\email{heinzmc@ornl.gov}
\affiliation{National Center for Computational Sciences, Oak Ridge National Laboratory, Oak Ridge, TN 37831, USA}
\affiliation{Physics Division, Oak Ridge National Laboratory, Oak Ridge, TN 37831, USA}
\affiliation{Technische Universit\"at Darmstadt, Department of Physics, 64289 Darmstadt, Germany}
\affiliation{ExtreMe Matter Institute EMMI, GSI Helmholtzzentrum f\"ur Schwerionenforschung GmbH, 64291 Darmstadt, Germany}
\affiliation{Max-Planck-Institut f\"ur Kernphysik, Saupfercheckweg 1, 69117 Heidelberg, Germany}

\author{A.~Schwenk \orcidlink{0000-0001-8027-4076}}
\email{schwenk@physik.tu-darmstadt.de}
\affiliation{Technische Universit\"at Darmstadt, Department of Physics, 64289 Darmstadt, Germany}
\affiliation{ExtreMe Matter Institute EMMI, GSI Helmholtzzentrum f\"ur Schwerionenforschung GmbH, 64291 Darmstadt, Germany}
\affiliation{Max-Planck-Institut f\"ur Kernphysik, Saupfercheckweg 1, 69117 Heidelberg, Germany}

\begin{abstract}

Low-resolution nuclear Hamiltonians, obtained from chiral effective field theory (EFT) and softened using renormalization group techniques, have been very successful in nuclear structure theory.
The associated EFT truncation uncertainty for these potentials is difficult to quantify.
We use singular value decompositions of low-resolution nuclear forces to obtain an operator basis to study Hamiltonian uncertainties for these potentials.
We perform Bayesian inference for the singular values and three-body low-energy constants, the free parameters of nuclear Hamiltonians in our framework, using likelihoods based on nucleon-nucleon phase shifts and triton observables to account for the EFT truncation uncertainties in these quantities.
Validating our inference, we find good reproduction of input uncertainties for low-energy phase shifts and three-nucleon observables.
On the other hand, uncertainties for higher-energy phase shifts are systematically underestimated,
which we attribute to limitations of the singular value decomposition and neglected correlations between phase shifts at different energies.
We propagate the resulting distribution of Hamiltonians forward to predictions for ground-state properties of $^{24,28}$O and $^{48}$Ca, comparing against other state-of-the-art nuclear structure predictions.
Our approach makes it possible to account for EFT uncertainties when using low-resolution potentials, which is important for many ongoing studies in exotic nuclei.

\end{abstract}

\maketitle

\section{Introduction}
\label{sec:intro}

Uncertainty quantification is a key aspect of modern nuclear theory. Theories of nuclear forces and most many-body methods are approximate, and the uncertainties associated with these approximations must be quantified. Chiral effective field theory (EFT) provides a systematically improvable description of nuclear forces rooted in quantum chromodynamics~\cite{Epelbaum2009RMP_ChiralEFTReview, Machleidt2011PR_ChiralEFTReview, Hammer2020RMP_NuclearEFT}.
Such forces have several sources of uncertainty, due to the truncation of the EFT, how the forces are regularized, and how they are fit to data.
Recent advances in Bayesian uncertainty quantification have enabled systematic approaches to quantifying theoretical uncertainties from various sources and propagating them to nuclear observables~\cite{Furnstahl2015JPGNPP_EFTBayesianUQ,Furnstahl2015PRC_EFTBayesianUQ, Epelbaum2015EPJA_EKMN3LO, Binder2016PRC_LENPICN4LONN, Carlsson2015PRX_UQNN, Melendez2017PRC_UQNN,Hu2022NP_Pb208, Kondo2023N_O28, Svensson2025arxiv}.

A notable gap in this body of work is uncertainty quantification for low-resolution Hamiltonians from chiral EFT~\cite{Bogner2010PPNP_RGReview}, which have been successful in providing accurate predictions for nuclear structure.
An interaction that stands out in the accurate description of binding energies and spectra across the nuclear chart is the \magic{} interaction~\cite{Hebeler2011PRC_SRG3NFits, Hagen2016PRL_Ni78, Simonis2017PRC_ChiralSaturationNuclei, Morris2018PRL_Sn100, Stroberg2021PRL_AbInitioLimits, Arthuis2024arxiv_LowResForces, Bonaiti2025arxiv}.
Despite this remarkable accuracy,
the uncertainties of this Hamiltonian are notoriously difficult to quantify.
Recent works have significantly advanced the quality of uncertainty estimates for nuclear structure observables~\cite{Hu2022NP_Pb208, Kondo2023N_O28}.
We aim to do the same for low-resolution Hamiltonians, combining their historical success with modern uncertainty quantification methods.

A key barrier to performing Bayesian inference and robust uncertainty quantification for low-resolution Hamiltonians is the lack of an explicit operator basis.
Low-resolution Hamiltonians are produced through nonlinear renormalization group (RG) transformations.
The RG transformations make them amenable to many-body calculations but destroy the underlying analytic structure of the potentials that is typically leveraged for uncertainty quantification.
While EFT truncation uncertainty estimates for such potentials are possible in some prescriptions~\cite{Huther2020N3LO, Maris2022PRC_LENPICSMSNuclei}, a complete treatment of uncertainties, going from few-body observables to Bayesian inference for low-energy constants (LECs) to posterior predictive distributions for nuclear structure~\cite{Hu2022NP_Pb208}, is still missing.

In this work, we overcome this barrier by leveraging data-driven singular value decompositions (SVDs) of nuclear forces.
Such decompositions have been demonstrated to precisely reproduce the input Hamiltonians using only a few terms,
meaning such decompositions effectively expose a low-rank operator basis for nuclear Hamiltonians~\cite{Tichai2021PLB_SVDNN, Zhu2021PRC_SVDNN, Tichai2022PRC_LSSVD, Tichai2023arxiv_SVD3N}.
Based on such decompositions, we are able to use state-of-the-art Bayesian uncertainty quantification approaches to account for EFT truncation uncertainties when using low-resolution Hamiltonians.
We apply these developments to study the structure of neutron-rich oxygen and calcium isotopes.

This paper is structured as follows:
In Sec.~\ref{sec:Method}, we describe the decomposition of low-resolution nucleon-nucleon (\nn{}) potentials and our selection of appropriate parameters. In Sec.~\ref{sec:Uncertainties}, we define our error models for \nn{} phase shifts and triton observables. We construct likelihoods based on the truncation uncertainties of these observables, perform Bayesian inference, sample the resulting posterior, and analyze posterior predictive distributions to check the consistency of our model.
In Sec.~\ref{sec:MediumMassApplications}, we apply the resulting posterior distributions to ground-state properties of $^{24,28}$O and $^{48}$Ca to investigate how EFT truncation uncertainties in few-body systems propagate to observables in medium-mass nuclei.
Finally, we conclude in Sec.~\ref{sec:Conclusion}.

\section{Linear operator structure}
\label{sec:Method}

Modern nuclear interactions derived from chiral EFT naturally exhibit a linear operator structure, where different contributions are summed up to compute the potential at a certain order~\cite{Epelbaum2009RMP_ChiralEFTReview, Machleidt2011PR_ChiralEFTReview}. For example, the leading order (LO) potential is
\begin{equation}
    V_{\mathrm{LO}} =  V_{\mathrm{ct}}^{(0)} +  V_{1\pi}^{(0)} = C_S + C_T\vec{\sigma}_1\cdot\vec{\sigma}_2 + V_{1\pi}^{(0)} \,,
\end{equation}
where $V_{1\pi}$ is the one-pion-exchange potential and $V_{\mathrm{ct}}^{(0)}$ are the LO contact interactions. $\vec{\sigma}_1$ and $\vec{\sigma}_2$ denote the nucleon spin operators, and $C_S$ and $C_T$ are LECs.
This allows us to write chiral EFT Hamiltonians as
\begin{equation}
    H(\mathbf{c}) = H_0 + \sum_i c_i H_i\,,
\end{equation}
where $H_0$ contains the parameter-free potential terms and the kinetic energy, and $c_i$ and $H_i$ are LECs and the associated operators in the parameter-dependent part of the Hamiltonian.

This linear structure facilitates a systematic statistical treatment of truncation uncertainties in chiral potentials.
It allows for the construction of efficient eigenvector continuation (EC) emulators for the calculation of observables~\cite{Frame2018PRL_EC, Konig2020PLB_UQEC, Duguet2023arxiv_EmulatorsReview}.
Eigenvector continuation emulators profit from a linear operator structure of the Hamiltonian by obtaining low-rank projections of the operators $H_0$ and $H_i$. The resulting EC matrix may then be quickly constructed and diagonalized as the LECs are varied.
Using such emulators, one can perform Bayesian inference for the LECs, arriving at a joint distribution for all LECs corresponding to a distribution of Hamiltonians~\cite{Wesolowski2016JPGNPP_BayesianInferenceEFT, Wesolowski2019JPGNPP_BayesianLECInferenceNN, Ekstrom2019PRL_SPCC,Jiang2022FP_ImportanceResamplingReview,Hu2022NP_Pb208, Kondo2023N_O28,Svensson2022PRC_BayesianInferenceHMC, Svensson2023PRC_NNBayesianInferenceN3LO, Svensson2023PRC_CorrelatedBayesNN_DeltaFull}.
To propagate this Hamiltonian uncertainty to other nuclear structure observables, one then samples Hamiltonians from this distribution, where the linear structure again makes it easy to construct the Hamiltonian for each sample: one simply sums up the operators $H_i$ together with the sampled $c_i$.

In practice, this linear operator structure is not always present for all chiral EFT interactions.
One of the most widely used interactions is the \magic{} interaction~\cite{Hebeler2011PRC_SRG3NFits}, which has been successfully applied to a range of nuclear structure calculations~\cite{Hagen2016PRL_Ni78, Simonis2017PRC_ChiralSaturationNuclei, Morris2018PRL_Sn100, Stroberg2021PRL_AbInitioLimits, Arthuis2024arxiv_LowResForces, Bonaiti2025arxiv}.
The potential is constructed from the N$^3$LO (next-to-next-to-next-to-leading order) \nn{} interaction by Entem and Machleidt (EM) with a $500\,\MeV{}$ cutoff \cite{Entem2003PRC_EM500}, evolved to $\lambda = 1.8\,\fminv{}$ through the similarity renormalization group (SRG)~\cite{Bogner2007PRC_SRGNN}, and an unevolved N$^2$LO three-nucleon (\nnn{}) interaction with a cutoff of $\Lambda_{\nnn} = 2\,\fminv{}$.
The \nnn{} LECs are fit to the \elem{H}{3} ground-state energy and \elem{He}{4} point-proton radius.
The SRG evolution of the \nn{} part improves convergence in many-body calculations~\cite{Bogner2008NPA_SRGNCSMConvergence, Bogner2010PPNP_RGReview}, but sacrifices the linear operator structure present in chiral EFT potentials~\cite{Bogner2007PRC_SRGNN}.
For conventional generator choices, the SRG flow equation is quadratic in the potential, meaning that different terms mix in nontrivial ways through the SRG evolution.
This leaves the evolved potential in a state where the original linear operator structure cannot be analytically recovered.

In this section, we aim to decompose SRG-evolved chiral potentials to recover a linear operator structure. In subsequent sections, we then leverage this decomposition for uncertainty quantification, the development and deployment of efficient emulators, and the easy evaluation of posterior predictive distributions for nuclear structure observables.

\subsection{Decomposition of the potential}

\begin{figure*}[t!]
    \centering
    \includegraphics[width=0.95\textwidth]{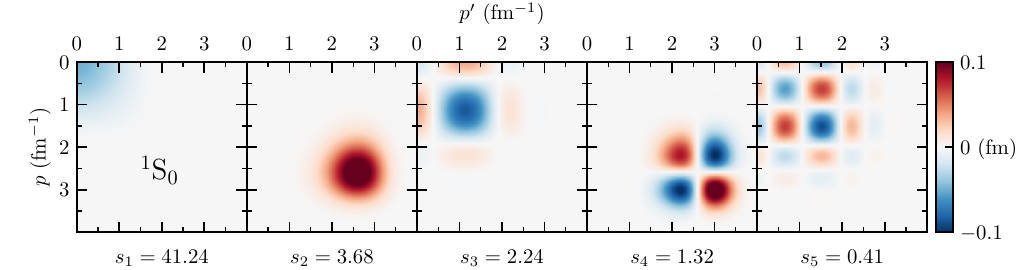}
    \includegraphics[width=0.95\textwidth]{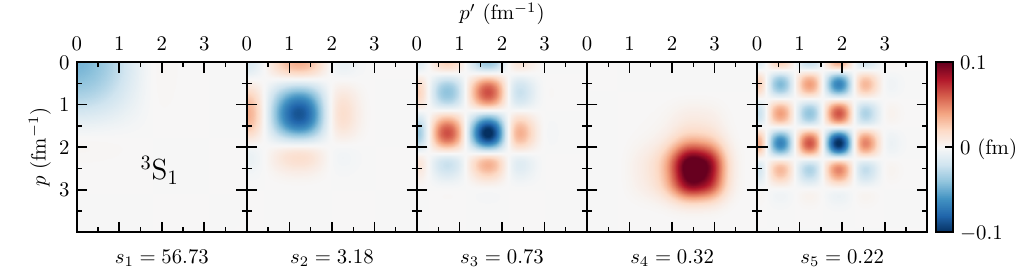}
    \caption{First five SVD operators for the N$^3$LO EM $500$ $\lambda=1.8\,\fminv{}$ potential in the $^1$S$_0$ (top panels) and $^3$S$_1$ (bottom panels) partial waves in momentum space representation $V(p, p')$. $s_i$ are the singular values of the corresponding operators.}
    \label{fig:SVD_basis}
\end{figure*}

We use the singular value decomposition to construct a new operator basis.
The resulting data-driven operators do not have a clear physics interpretation, but provide a quantitative reproduction of \nn{} potentials using only a few leading terms~\cite{Tichai2021PLB_SVDNN}.

The SVD decomposes any $m\times n$ matrix $V$ into the product of three matrices,
\begin{equation}
    V = LS R^{\dagger} \,,
    \label{SVD_1}
\end{equation}
with the left and right matrices $L$ and $R^{\dagger}$ and the singular value matrix $S = \mathrm{diag}(s_i)$, where $s_i$ are the nonnegative singular values ordered with decreasing magnitude. 
In general, $L$ and $R^{\dagger}$ are $m\times m$ and  $n\times n$ unitary matrices, respectively. They are made up of the left and right singular vectors, the outer product of which forms the individual SVD operators. $S$ has dimensions $m\times n$ and consists of nonnegative real values on its diagonal.
When decomposing square $n\times n$ matrices, like for our $V_{\mathrm{NN}}$ potentials, the dimensions of $L$, $R^{\dagger}$, and $S$ are also $n \times n$.

Rewriting Eq.~\eqref{SVD_1} yields
\begin{equation}
    V = \sum_i s_i \ket{L_i}\bra{R_i} \,,
\end{equation}
which clearly has a linear operator structure with the coefficients $s_i$ and the associated operators $O_i$ constructed as outer products of left and right vectors $\ket{L_i}\bra{R_i}$.
It is this linear operator structure that we leverage in this work. We decompose \nn{} potentials in a momentum-space partial-wave basis,
\begin{equation}
V(p,p') = \langle p \, (l S) J T M_T | V_\nn | p' \, (l' S) J T M_T \rangle\,,
\end{equation}
with the final and initial orbital angular momenta  $l$ and $l'$, the two-body spin $S$, the two-body total angular momentum $J$, the two-body isospin $T$ with projection $M_T$, and the absolute values of the outgoing and incoming relative momenta $p$ and $p'$.
The matrix $V(p,p')$ is decomposed in each partial wave,
automatically preserving relevant symmetries of the potential, e.g., rotational invariance.

In Fig.~\ref{fig:SVD_basis}, we show the first five SVD operators and associated singular values for the $^1$S$_0$ and $^3$S$_1$ partial waves of the N$^3$LO EM $500$ $\lambda=1.8\,\fminv{}$ interaction.
In both partial waves the first operator is purely attractive. 
In both cases, we also identify a purely  repulsive operator with strength at high momenta, $O_2$ in $^1$S$_0$ and $O_4$ in $^3$S$_1$.
For higher $s_i$ the operators start showing oscillatory behavior.
For both partial waves, $s_5$ is at least two orders of magnitude smaller than $s_1$.
This hierarchy of singular values allows us to use SVD-decomposed potentials truncated at rank five
\begin{equation}
    \tilde{V} =\sum_{i=1}^{R_{\mathrm{SVD}}=5}s_i\ket{L_i}\bra{R_i},
\end{equation}
where $R_{\mathrm{SVD}}$ denotes the rank of the decomposition.
It was shown in \cite{Tichai2021PLB_SVDNN} that the first five out of the $100$ operators of a decomposed potential are sufficient to reach accuracies above $99\%$ for various observables.
We are now left with SRG-evolved interactions in a linear operator structure for any partial wave, each depending on five parameters.
We treat the parameters $s_i$ as uncertain and seek to perform Bayesian inference for their values.

In order to better understand the impact of the operators, we individually vary the singular values of a partial-wave decomposed potential in an interval of $[0.5s_i, 1.5s_i]$. Afterwards, we compute the resulting phase shifts to understand which energies are affected by the individual operators. We also use this to identify operators that do not significantly affect phase shifts at the energies we are interested in. Those operators will not be varied in the future treatment of the potentials. This is done in order to reduce the number of parameters in our framework.

\begin{table}[t!]
\centering
\begin{tabular}{c|cccccc}
               & $^1$S$_0$ & $^3$S$_1$ & $^1$P$_1$ & $^3$P$_0$ & $^3$P$_1$ & $^3$P$_2$ \\ \hline
$s_1$          & x       & x       & x       & x       & x       & x       \\
$s_2$          & –       & x       & x       & –       & –       & x       \\
$s_3$          & x       & x       & x       & –       & x       & x       \\
$s_4$          & –       & –       & –       & x       & x       & –       \\
$s_5$          & x       & –       & –       & –       & –       & –       \\
\end{tabular}
\caption{Selection of singular values $s_i$ for uncertainty propagation. Singular values that are subsequently varied for uncertainty quantification are marked with an ``x'', while ``-'' indicates fixed singular values.}
\label{tab:singular_values}
\end{table}

\begin{figure*}[t!]
    \centering
    \includegraphics[width=\textwidth]{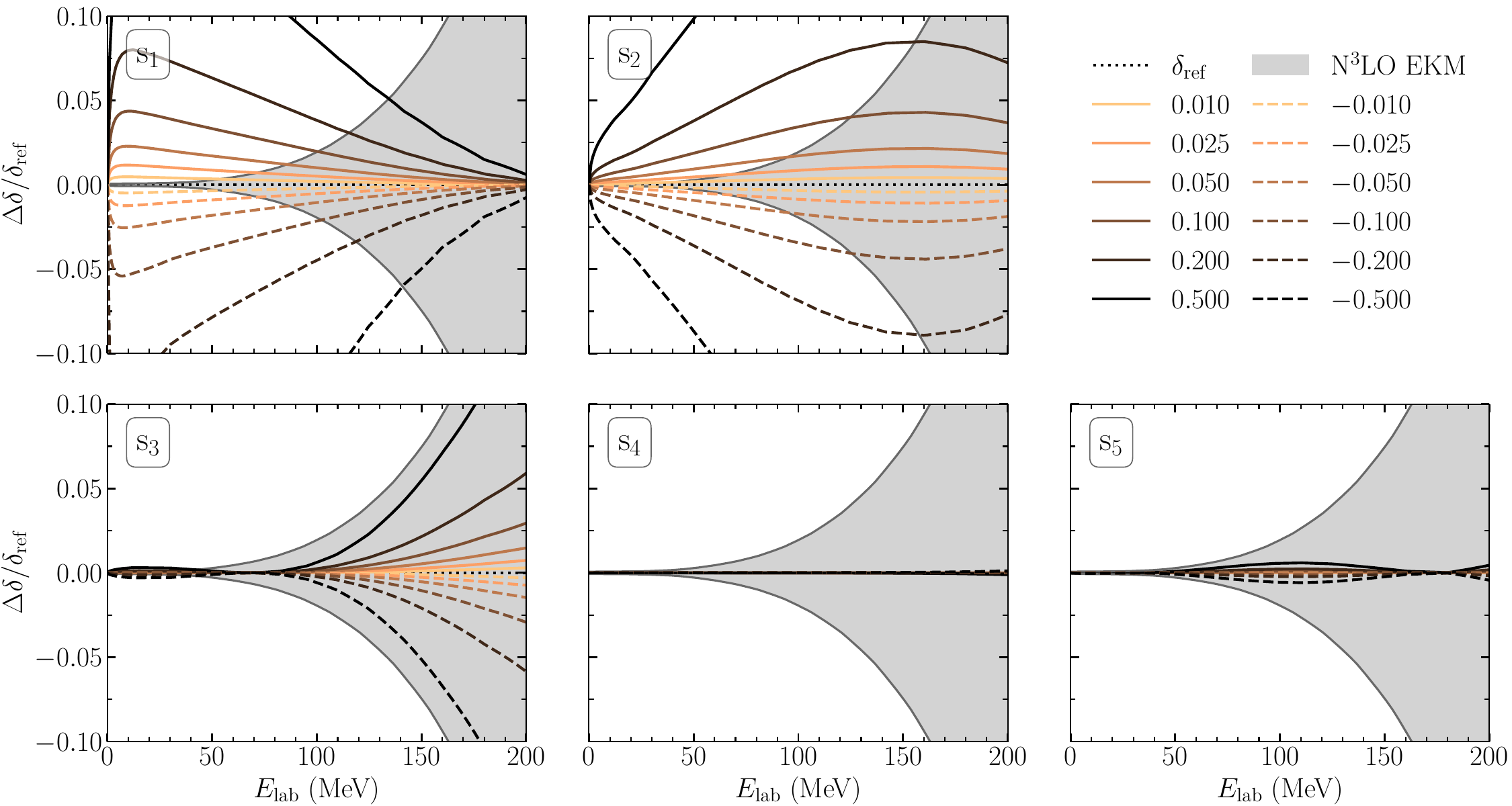}
    \caption{Relative phase shifts for singular value variations in the $^3$S$_1$ partial wave. Each panel shows the change of phase shifts when varying one singular value at a time. Different colors represent varied singular values by a factor as given in the legend, where $0.5$ stands for a singular value increased by $50\%$. Solid lines indicate positive factors while dashed lines stand for negative factors.
    The reference value is obtained by using the SVD-decomposed N$^3$LO EM $500$ $\lambda=1.8\,\fminv{}$ interaction at rank five with the default set of singular values to solve the Lippmann-Schwinger equation. The gray band represents the relative EKM truncation uncertainty at N$^3$LO.}
    \label{fig:SVD_phase_shifts}
\end{figure*}

Figure \ref{fig:SVD_phase_shifts} shows the resulting change in the phase shifts in the $^3$S$_1$ partial wave up to $200\,\MeV{}$ relative to the reference $\delta_{\mathrm{ref}}$, which is the result of solving the Lippmann-Schwinger equation with unvaried singular values. The gray band represents the EFT truncation uncertainty at N$^3$LO [following the EKM prescription as defined in Eq.~\eqref{eq:EKM}].  Each panel shows the effects of varying a single singular value by various percentages.
We easily identify that the fourth and fifth operators of this potential have negligible impact on phase shifts at energies below $200\,\MeV$. We still keep these operators in the potential, but keep their respective singular values fixed. 

Operators that barely affect phase shifts up to $200\,\MeV{}$ can be identified by their large high-momentum matrix elements. This can be seen in Fig.~\ref{fig:SVD_basis}. The fourth and fifth operators have their largest matrix elements at momenta above $1.5\,\fminv{}$, corresponding to a laboratory energy of $186\,\MeV{}$.
Looking at the three remaining operators, we observe that each of the operators appears to roughly affect phase shifts at different energies. This is important, as we aim to constrain the singular values through phase shifts at a range of energies up to $200\,\MeV{}$. If all operators were to yield similar variations in phase shifts, we would not expect to be able to independently constrain our parameters and fully capture the EFT truncation uncertainties by calibrating to phase shift data.

In Table~\ref{tab:singular_values}, we identify three singular values to be varied to capture phase shift uncertainties in each partial wave with the exception of $^3$P$_0$, where we only find two appropriate singular values. This leaves us with $17$ NN parameters in total. Note that we do not include higher order SVD operators beyond $R_\mathrm{SVD}=5$ for these partial waves, so subsequent results are subject to small SVD truncation errors, which we quantify. All other partial wave interactions are not SVD decomposed.
With a parameter-dependent description of our potentials at hand, we can move on to performing statistical inference for these parameters.

\section{EFT truncation uncertainties}
\label{sec:Uncertainties}

A successful approach to quantifying EFT uncertainties for observables assumes a power series expansion of the observable of interest $y$~\cite{Furnstahl2015PRC_EFTBayesianUQ},
\begin{equation}
y = y_{\mathrm{ref}}\sum_{n=0}^{\infty} c_n \left(\frac{Q}{\Lambda_b}\right)^n,
\label{power_series_expansion}
\end{equation}
with the dimensionless coefficients $c_n$, the natural size of the observable $y_{\mathrm{ref}}$, and the expansion parameter $Q/\Lambda_b$.
This parameter consists of the momentum scale $Q$ and the EFT breakdown scale $\Lambda_b$.
For NN scattering, the relevant momentum scale is set by $Q\equiv \max \left( p, m_{\pi}\right)$, where $p$ is the relative momentum between two nucleons in the center-of-mass frame and $m_\pi$ is the pion mass.

Truncating the EFT at finite order $k$ also truncates this expansion,
\begin{equation}
y_k = y_{\mathrm{ref}}\sum_{n=0}^{k} c_n \left(\frac{Q}{\Lambda_b}\right)^n.
\label{power_series_expansion_finite}
\end{equation}
The resulting error due to this truncation is simply the sum over all higher-order neglected terms:
\begin{equation}
\Delta y_k \equiv y_{\mathrm{ref}}\sum_{n=k+1}^{\infty} c_n \left(\frac{Q}{\Lambda_b}\right)^n.
\label{ew:uncertainty_power_series}
\end{equation}
These higher-order corrections are generally unavailable to us, so we estimate the truncation uncertainty by the first omitted term $\Delta y_k \approx c_{k+1}(Q/\Lambda_b)^{k+1}$.
To estimate the size of $c_{n+1}$, we follow the prescription of Epelbaum, Krebs, and Meißner (EKM) \cite{Epelbaum2015EPJA_EKMN3LO}, assuming the coefficients $c_n$ to be of natural size. The resulting uncertainty at $ \mathrm{N^3LO}$ is then given by
\begin{align}
\Delta y^{\mathrm{N^3LO}}_{\mathrm{EKM}} & = \max \left( \left(\frac{Q}{\Lambda_b}\right)^{5} |y^{\mathrm{LO}}|, \right.
\left. \left(\frac{Q}{\Lambda_b}\right)^{3} |y^{\mathrm{LO}} - y^{\mathrm{NLO}}|, \right.\nonumber\\
&\quad\hphantom{\max \Bigg(} \left. \left(\frac{Q}{\Lambda_b}\right)^{2} |y^{\mathrm{NLO}} - y^{\mathrm{N^2LO}}|, \right. \nonumber \\
 &\quad\hphantom{\max \Bigg(} \left. \left(\frac{Q}{\Lambda_b}\right) |y^{\mathrm{N^2LO}} - y^{\mathrm{N^3LO}}| \right).
\label{eq:EKM}
\end{align}

\begin{table*}[t!]
\centering
\caption{Absolute phase-shift uncertainties (in degree) at N$^3$LO for S- and P-waves of the EMN $500$ $\lambda=2.0\,\text{fm}^{-1}$ interaction at different energies. Uncertainties are obtained through the EKM prescription.}
\label{table:phase_shift_uncertainties}
\renewcommand{\arraystretch}{1.3}
\begin{ruledtabular}
\begin{tabular}{c c c c c c c}
$E_\mathrm{lab}$ (MeV) & $\Delta\delta_{^1\mathrm{S}_0}$ & $\Delta\delta_{^3\mathrm{S}_1}$ & $\Delta\delta_{^1\mathrm{P}_1}$ & $\Delta\delta_{^3\mathrm{P}_0}$ & $\Delta\delta_{^3\mathrm{P}_1}$ & $\Delta\delta_{^3\mathrm{P}_2}$ \\ \midrule
$\hphantom{00}1$   & $4.38 \times 10^{-2}$ & $1.01 \times 10^{-1}$ & $5.45 \times 10^{-4}$ & $8.91 \times 10^{-4}$ & $2.99 \times 10^{-4}$ & $6.47 \times 10^{-4}$ \\ 
$\hphantom{00}5$   & $7.01 \times 10^{-2}$ & $8.22 \times 10^{-2}$ & $5.60 \times 10^{-3}$ & $1.08 \times 10^{-2}$ & $2.93 \times 10^{-3}$ & $6.46 \times 10^{-3}$ \\ 
$\hphantom{0}10$  & $1.04 \times 10^{-1}$ & $7.24 \times 10^{-2}$ & $1.40 \times 10^{-2}$ & $3.31 \times 10^{-2}$ & $7.11 \times 10^{-3}$ & $1.56 \times 10^{-2}$ \\ 
$\hphantom{0}20$  & $1.55 \times 10^{-1}$ & $6.71 \times 10^{-2}$ & $2.98 \times 10^{-2}$ & $9.32 \times 10^{-2}$ & $1.49 \times 10^{-2}$ & $3.04 \times 10^{-2}$ \\ 
$\hphantom{0}30$  & $1.97 \times 10^{-1}$ & $8.61 \times 10^{-2}$ & $3.95 \times 10^{-2}$ & $1.54 \times 10^{-1}$ & $2.46 \times 10^{-2}$ & $3.48 \times 10^{-2}$ \\ 
$\hphantom{0}50$  & $3.54 \times 10^{-1}$ & $1.76 \times 10^{-1}$ & $4.81 \times 10^{-2}$ & $3.42 \times 10^{-1}$ & $4.60 \times 10^{-2}$ & $6.70 \times 10^{-2}$ \\ 
$\hphantom{0}75$  & $8.28 \times 10^{-1}$ & $4.21 \times 10^{-1}$ & $1.01 \times 10^{-1}$ & $8.51 \times 10^{-1}$ & $5.99 \times 10^{-2}$ & $2.21 \times 10^{-1}$ \\ 
$100$ & $1.51 \times 10^{0\hphantom{-}}$  & $8.35 \times 10^{-1}$  & $2.30 \times 10^{-1}$ & $1.53 \times 10^{0\hphantom{-}}$  & $1.61 \times 10^{-1}$ & $4.94 \times 10^{-1}$ \\
$150$ & $3.53 \times 10^{0\hphantom{-}}$  & $2.25 \times 10^{0\hphantom{-}}$  & $7.40 \times 10^{-1}$ & $3.27 \times 10^{0\hphantom{-}}$  & $7.66 \times 10^{-1}$ & $1.39 \times 10^{0\hphantom{-}}$  \\ 
$200$ & $6.46 \times 10^{0\hphantom{-}}$  & $4.42 \times 10^{0\hphantom{-}}$  & $1.65 \times 10^{0\hphantom{-}}$  & $5.43 \times 10^{0\hphantom{-}}$  & $1.73 \times 10^{0\hphantom{-}}$  & $2.58 \times 10^{0\hphantom{-}}$  \\ 
\end{tabular}
\end{ruledtabular}
\end{table*}

When using the \magic{} interaction, applying this uncertainty quantification prescription to nuclear structure observables is challenging for several reasons.
First, the \magic{} interaction is constructed from \nn{} and \nnn{} interactions at different orders. The \nn{} interaction is at N$^3$LO in the chiral expansion, while the \nnn{} interaction is at N$^2$LO. 
Second, this approach requires a definition of the momentum scale $Q$, which is difficult to define unambiguously for nuclei.
It is, however, possible to infer the dimensionless expansion parameter $Q/\Lambda_b$ using Bayesian methods~\cite{ Drischler2020PRC_NuclearMatterGPBLong, Wesolowski2021PRC_3NInference,Ekstrom2024PLB_PionlessEFTBreakdownScale}.

To address these challenges, we use EFT uncertainties for observables in two- and three-body systems to construct a likelihood for our theory predictions.
We perform Bayesian inference based on this likelihood to obtain a joint posterior distribution for the free parameters $\boldsymbol{\alpha}_{\nn{}}$, $c_D$, and $c_E$ in our \nn{} and \nnn{} potentials, where $\boldsymbol{\alpha}_{\nn{}}$ denotes all \nn{} parameters and $c_D$ and $c_E$ are 3N LECs.
Sampling from this distribution yields an ensemble of nuclear Hamiltonians that we can use in nuclear structure calculations.
This allows us to propagate the EFT uncertainty captured in the posterior distribution to nuclear structure calculations.
In the following, we introduce the construction of our NN likelihood based on NN scattering phase shifts, the construction of our 3N likelihood based on triton observables, the choices of priors for $\boldsymbol{\alpha}_{\nn{}}$, $c_D$, and $c_E$, and the joint inference based on these likelihoods and priors.

\subsection{NN uncertainties}

We begin by constructing a likelihood function for the uncertainties in the NN interactions.
In order to infer a distribution of the singular values, we incorporate data with Bayes' theorem. Specifically, for each partial wave, we infer distributions for $s_i$ conditioned on phase shifts and their corresponding EFT truncation uncertainties.

Bayes' theorem allows us to capture truncation uncertainties of phase shifts while also implementing prior knowledge of the singular value distributions:
\begin{equation}   
\text{pr}(\boldsymbol{\alpha}|\mathcal{D}) \propto  \mathcal{L}(\boldsymbol{\alpha})\text{pr}(\boldsymbol{\alpha})=\text{pr}(\mathcal{D}|\boldsymbol{\alpha})\text{pr}(\boldsymbol{\alpha}) \,,
\end{equation}
where $\text{pr}(\boldsymbol{\alpha}|\mathcal{D})$ is the posterior probability of the set of singular values in a single partial wave given a set of data $\mathcal{D}$, which in our case are phase shifts~\cite{Jiang2022FP_ImportanceResamplingReview}. $\mathcal{L}(\boldsymbol{\alpha}) \equiv \text{pr}(\mathcal{D}|\boldsymbol{\alpha}) $ is the likelihood. It expresses the conditional probability of $\mathcal{D}$ given the parameters $\boldsymbol{\alpha}$. In our case, $\boldsymbol{\alpha}$ are the three singular values that parametrize the \nn{} potential in a single partial wave.
$\text{pr}(\boldsymbol{\alpha})$ is the prior where we can implement prior knowledge or assumptions about the distribution of singular values.
We use a uniform prior distribution on $\boldsymbol{\alpha}$ with no lower and upper bounds.
This amounts to sampling directly from the likelihood,
which is a reasonable choice given that the likelihood is strongly constraining for the leading singular values in each partial wave (see Fig.~\ref{fig:SVD_phase_shifts}).
We also performed a second inference using a scale-invariant prior to check prior sensitivity,
and we discuss this with the results of the inference below.

Our goal is now to construct a likelihood function from \nn{} phase shifts incorporating the EFT truncation uncertainties. 
By calculating the EKM uncertainties as defined in Eq.~\eqref{eq:EKM}, with $\Lambda_b = 600\,\MeV$ as suggested in~\cite{Epelbaum2015EPJA_EKMN3LO}, we obtain an interval around the reference phase shift at a given energy, $\Delta\delta_{\mathrm{EKM}}(E)$.
To assign probabilities at a given energy, we assume a Gaussian distribution for the EKM uncertainties with the standard deviation $\Delta\delta_{\mathrm{EKM}}(E)$.
The expectation value for this Gaussian is the reference phase shift obtained from using the default set of singular values.

Note that we need order-by-order results to assess uncertainties this way. However, the EM interaction is only available at N$^3$LO. We use the more recent EMN interactions~\cite{Entem2017PRC_ChiralEMN}, which are available up to N$^4$LO to calculate phase shifts at different chiral orders.
We evolve these interactions to $\lambda=2.0\,\fminv$. The EMN interactions have the same cutoff of $500\,\MeV$ and the same regularization scheme, making them appropriate for estimating EFT uncertainties for the N$^3$LO EM interaction.
The resulting truncation uncertainties for S- and P-wave phase shifts at different energies up to $200\,\MeV{}$ are given in Table~\ref{table:phase_shift_uncertainties}.
For each partial wave and for each energy $E$, we end up with a Gaussian likelihood 
\begin{equation}
    f(\boldsymbol{\alpha}, E) = \frac{1}{\Delta\delta_{\mathrm{EKM}}(E) \sqrt{2\pi}} \exp\left(-\frac{\left(\delta_{\boldsymbol{\alpha}}(E) - \delta_{\mathrm{ref}}(E)\right)^2}{2[\Delta\delta_{\mathrm{EKM}}(E)]^2}\right),
\end{equation}
based on the predicted phase shift $\delta_{\boldsymbol{\alpha}}(E)$ and the reference phase shift $\delta_{\mathrm{ref}}(E)$.

\begin{figure}[t!]
    \centering
    \includegraphics[width=1\linewidth]{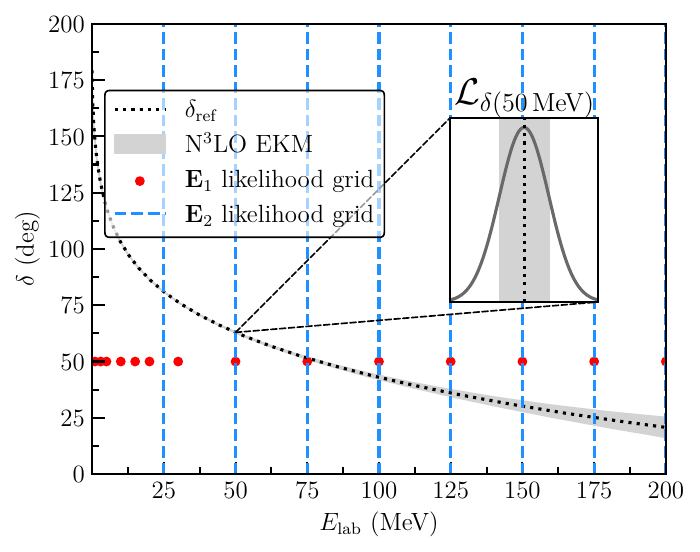}
    \caption{Likelihood construction for phase shifts in a single partial wave. The dotted black line represents the reference phase shifts in the $^3$S$_1$ partial wave obtained by using the EM $500$ $\lambda=1.8\,\fminv$ interaction at N$^3$LO, SVD decomposed and truncated at rank five. The gray band indicates the EKM uncertainty at N$^3$LO. Gaussian likelihoods are constructed at different energies according to the red points ($\mathbf{E}_1$) or the blue dashed lines ($\mathbf{E}_2$), defined in Eqs.~\eqref{eq:E1} and \eqref{eq:E2}.}
    \label{fig:NN_likelihood}
\end{figure}

As phase shifts are given as a function of energy, we want to evaluate our likelihood at multiple energies as well.
We construct a multivariate Gaussian for $n$ different energies in the range between $0\,$MeV and $200\,$MeV. The upper bound of the interval is an assumption, realizing our focus on low energies.
In Fig.~\ref{fig:NN_likelihood}, we visualize the construction of the \nn{} likelihood for two different energy grids $\bf{E}_1$ and $\bf{E}_2$ defined in Sec.~\ref{sec:PosteriorSampling}.
The likelihood is then given by
\begin{align}
    &\mathcal{L}( \boldsymbol{\alpha}) = \frac{1}{\sqrt{(2\pi)^n \mathrm{det}(\Sigma)}} \times\nonumber \\
    &\exp\left(-\frac{1}{2} \left[\boldsymbol{\boldsymbol{\delta}}_{\boldsymbol{\alpha}}(E) - \boldsymbol{\delta}_{\mathrm{ref}}(E)\right]^T \Sigma^{-1} \left[\boldsymbol{\boldsymbol{\delta}}_{\boldsymbol{\alpha}}(E) - \boldsymbol{\delta}_{\mathrm{ref}}(E)\right]\right),
    \label{eq:likelihood}
\end{align}
with the expectation value vector
\begin{equation}
\boldsymbol{\delta}_{\mathrm{ref}}(E) = (\delta_{\mathrm{ref}}(E^{(1)}), ..., \delta_{\mathrm{ref}}(E^{(n)})) \,,
\end{equation}
and the singular-value-dependent phase shift result
\begin{equation}
\boldsymbol{\delta}_{\boldsymbol{\alpha}}(E) = (\delta_{\boldsymbol{\alpha}}(E^{(1)}), ..., \delta_{\boldsymbol{\alpha}}(E^{(n)}))\,.
\end{equation}
Both are $n$-dimensional vectors. Their size is determined by the amount of energy points in our likelihood.
The $n \times n$ covariance matrix is denoted as $\Sigma$.
We assume in the definition of the covariance matrix $\Sigma$ that there is zero correlation between the phase shifts at different energies
\begin{equation}
    \Sigma = \begin{bmatrix}
[\Delta\delta_{\mathrm{EKM}}(E^{(1)})]^2  &   &0  \\
 & \ddots &   \\
0 &  & [\Delta\delta_{\mathrm{EKM}}(E^{(n)})]^2 \\
\end{bmatrix}.
\label{eq:cov_matrix_NN}
\end{equation}
In the future, we aim for a correlated treatment of phase shift uncertainties through the use of Gaussian processes~\cite{Melendez2019PRC_CorrelatedEFTUQ}.

We construct the combined likelihood for our NN potentials combining the individual likelihoods for the six partial waves
\begin{equation}
\begin{aligned}
    \mathcal{L}_{\mathrm{NN}}(\boldsymbol{\alpha}_{\mathrm{NN}}) =\,&
    \mathcal{L}_{^{^1\mathrm{S}_0}}(\boldsymbol{\alpha}_{^{^1\mathrm{S}_0}})
    \mathcal{L}_{^{^3\mathrm{S}_1}}(\boldsymbol{\alpha}_{^{^3\mathrm{S}_1}})
    \mathcal{L}_{^{^1\mathrm{P}_1}}(\boldsymbol{\alpha}_{^{^1\mathrm{P}_1}}) \\
    &\times \mathcal{L}_{^{^3\mathrm{P}_0}}(\boldsymbol{\alpha}_{^{^3\mathrm{P}_0}})
    \mathcal{L}_{^{^3\mathrm{P}_1}}(\boldsymbol{\alpha}_{^{^3\mathrm{P}_1}})
    \mathcal{L}_{^{^3\mathrm{P}_2}}(\boldsymbol{\alpha}_{^{^3\mathrm{P}_2}}) \,,
\end{aligned}
\label{eq:likelihood_NN_6}
\end{equation}
by multiplying the individual likelihoods for each partial wave. The singular values from all S- and P-waves are collectively denoted  $\boldsymbol{\alpha}_{\mathrm{NN}}$.
In doing so, we assume zero correlation between individual partial waves.
The resulting likelihood $ \mathcal{L}_{\mathrm{NN}}$ is a multivariate Gaussian with dimension $d = 6\times \mathrm{dim}(\mathbf{E})$ for the chosen energy grid $\mathbf{E}$.

The main reason we only consider S- and P-waves is to limit the number of parameters in our model. 
The three parameters we assigned to each partial wave do not take into account different total isospin projections. Out of the six considered partial waves, only $^3$S$_1$ and $^1$P$_1$ appear in an isospin singlet. 
For partial waves $^1$S$_0$, $^3$P$_0$, $^3$P$_1$, and $^3$P$_2$, the isospin wave function must be symmetric, implying a total isospin quantum number $T = 1$ corresponding to a triplet state. 
This results in three distinct interactions within each partial wave, corresponding to the three projections of the total isospin.
Since we determine three parameters to vary each potential, we would end up with $4\times3\times3 + 2\times3\times1 = 42$ parameters in total. We assume charge independence to reduce the number of parameters,
\begin{equation}
V_{nn} \approx V_{np} \approx V_{pp} - V_C \,,
\label{eq:charge_independence}
\end{equation}
with the Coulomb interaction $V_C$.
This is realized by replacing both $V_{nn}$ and $V_{pp} - V_C$ with $V_{np}$. 
After modifying the $pp$ interaction through the charge independence assumption and the SVD, we add the unmodified Coulomb interaction to it.
Now all interactions inside an isospin triplet depend on the same three parameters, reducing the number of parameters to $4\times3\times1 + 2\times3\times1 = 18$.
 We assess the validity of this approximation in Sec.~\ref{sec:MediumMassApplications}.
Note that we only use $17$ NN parameters, as we limit ourselves to varying two parameters in the $^3$P$_0$ partial wave where only two of the largest five singular values have a significant impact on phase shift variations over these energies.

\subsection{\nnn{} uncertainties}

Three-nucleon forces first appear at N$^2$LO in chiral EFT.
These interactions are parametrized by five LECs $c_1$, $c_3$, $c_4$, $c_D$, and $c_E$, where $c_1$, $c_3$, and $c_4$ also appear in the NN interactions.
Similarly to the two-nucleon interactions, we need to find a way to constrain these parameters.
We limit ourselves to varying the two short-range LECs $c_D$ and $c_E$ and leave the long-range LECs $c_i$ unchanged.
Their values are fixed to $c_1 = -0.81\,\text{GeV}^{-1}$, $c_3 = -3.2\,\text{GeV}^{-1}$, and $c_4 = 5.4\,\text{GeV}^{-1}$~\cite{Entem2003PRC_EM500}.

To infer distributions for the \nnn{} LECs, we follow the work by Wesolowski {\it et al.}~\cite{Wesolowski2021PRC_3NInference}.
We choose the triton ground-state energy $E(^3\text{H})$ as well as the comparative triton half-life $fT_{1/2}$~\cite{Gazit2008PRL_3NLECS} as additional observables for our full likelihood.
These both depend on $c_D$ and $c_E$, but also on all the \nn{} singular values $\boldsymbol{\alpha}_{\nn{}}$.
We compute these observables using EC-based no-core shell model emulators~\cite{Konig2020PLB_UQEC} constructed by Takayuki Miyagi~\cite{takayuki_private}. 

In order to construct a likelihood from these observables, we again start with Eq.~\eqref{power_series_expansion}, as we are interested in the EFT truncation error.
However, this time we will not use the EKM uncertainties.
Assuming the series coefficients $c_n$ to be of natural size,
we can use a constant $\bar{c}$, move it out of the power series, and write down the geometric series expression
\begin{equation}
    \left( \Sigma \right)_{ij} = \left[\frac{(y_{\text{EM}} \bar{c} (\frac{Q}{\Lambda_b})^{k+1})^2}{1 - (\frac{Q}{\Lambda_b})^2}\right] \delta_{ij} \,,
    \label{eq:3N_cov}
\end{equation}
where we use the \magic{} prediction $y_{\text{EM}}$ as the characteristic size of the observable $y$~\cite{Melendez2017PRC_UQNN, Furnstahl2015PRC_EFTBayesianUQ}.
This gives us a diagonal covariance matrix as we do not assume any correlation between the observables.
Since we use a single covariance matrix for multiple observables,
the variables $\tilde{Q}\equiv Q/\Lambda_b$ and $\bar{c}$ are determined globally for both of our observables.
This was done in~\cite{Wesolowski2021PRC_3NInference} by investigating the NLO to N$^2$LO corrections of triton and helium energies and radii and treating $\tilde{Q}$ and $\bar{c}$ as random variables.
This results in a distribution for the two variables, we use their central values of approximately $\tilde{Q} = 1/3 $ and $\bar{c} = 1$.

We note that the triton comparative half-life was not included in the inference of $\tilde{Q}$ and $\bar{c}$, as its order-by-order convergence was found to be poorly behaved~\cite{Wesolowski2021PRC_3NInference}.
Recent work explains this behavior by the symmetry protection of the Gamow-Teller matrix elements~\cite{LiMuli2025arxiv_GammowTellerSymmetry}. 
This causes the actual corrections to the leading-order predictions to be unexpectedly small when compared to those of radii and energies.
While the triton comparative half-life still serves as an effective constraint for the LEC $c_D$ the assessment of uncertainties through EFT truncations of radii and energies as in \cite{Wesolowski2021PRC_3NInference} is likely to overestimate the true uncertainties.

Using this covariance matrix, we construct a multivariate Gaussian as our $3$N likelihood,
\begin{equation}
    \mathcal{L}_{\nnn{}}(c_D, c_E,\boldsymbol{ \alpha}_{\mathrm{NN}}) = \mathcal{N}(\boldsymbol{y}_{\text{ref}}, \Sigma) \,,
    \label{eq:likelihood_3N}
\end{equation}
with the reference theoretical predictions from our emulators $\boldsymbol{y}_{\text{ref}} = (E(^3\text{H})_{\text{ref}}, fT_{1/2_{\text{ref}}})$ for a set of default parameters. For the \nn{} interaction, we choose the unvaried singular values. For the \nnn{} interaction, we use the fixed $c_i$ values and $c_D = 1.264$ and $c_E = -0.12$ to calculate $y_{\mathrm{ref}}$.
Note that the likelihood also depends on the NN parameters $\boldsymbol{ \alpha}_{\mathrm{NN}}$.

With Bayes' theorem we arrive at an expression for the distribution of the LECs,
\begin{align}
&\text{pr}(c_D, c_E, \boldsymbol{\alpha}_{\mathrm{NN}} | E(^3\text{H}), fT_{1/2})
\nonumber \\
&\quad \propto \text{pr}(E(^3\text{H}), fT_{1/2} | c_D, c_E, \boldsymbol{\alpha}_{\mathrm{NN}}) \nonumber \\
&\qquad \times
\text{pr}(c_D, c_E, \boldsymbol{\alpha}_{\mathrm{NN}}) \,,
\end{align}
with the prior $\text{pr}(c_D, c_E,\boldsymbol{ \alpha}_{\mathrm{NN}})$ consisting of uniform distributions for $c_D, c_E$ around their central values, and the uniform priors for the singular values.
The parameter $c_D$ has a prior range of $[-8.7, 11.3]$ and $c_E$ has a range of $[-2.1, 1.9]$. These intervals have shown no noticeable effects on the posterior distribution.
The likelihood $\mathcal{L}(c_D, c_E,\boldsymbol{ \alpha}_{\mathrm{NN}}) \equiv\text{pr}(E(^3\text{H}), fT_{1/2}|c_D, c_E,\boldsymbol{ \alpha}_{\mathrm{NN}})$ is conditioned on the truncation uncertainty of the triton binding energy and comparative half-life.

The comparative half-life of the triton is an electroweak observable, related to transition matrix elements by
\begin{equation}
    fT_{1/2} = \frac{K/G_V^2}{\braket{\hat{F}}^2 + \frac{f_A}{f_V}g_A^2\braket{\hat{GT}}^2},
\end{equation}
with $K/G_V^2 = 6147$, $f_A/f_V \approx 1$, and the axial-vector coupling constant $g_A = 1.27$ \cite{Schiavilla1998PRC_TritonHalfLife}.
$\braket{\hat{F}}$ and $\braket{\hat{GT}}$ denote the reduced Fermi and Gamow-Teller matrix elements. We compute these using emulated ground-state wave functions.
The Gamow-Teller transition involves the axial-vector current and is sensitive to two-body currents, which depend on the three-body LEC $c_D$, so it provides an effective tool to constrain $3$N interactions~\cite{Gazit2008PRL_3NLECS}.

Following Eq.~\eqref{eq:3N_cov}, we obtain the standard deviations shown in Table~\ref{tab:3N_oberservables}.
We denote the results using the \magic{} interaction as $y_{\text{EM}}$ and those using our emulators with default parameter values as $y_{\text{ref}}$ in Table~\ref{tab:3N_oberservables}.
We notice that the emulator predictions deviate from the $1.8/2.0$~(EM) results for the triton energy. The emulator uncertainty is below $1\%$ \cite{takayuki_private}. The deviations between the emulator and the \magic{} predictions are primarily caused by the charge-independence assumption and the SVD truncation.

\begin{table}[t!]
    \centering
    \caption{Values denoted as $y_{\text{ref}}$ are the results calculated with the emulator when using the default set of parameters. Deviations from the results obtained using the $1.8/2.0$ (EM) interaction without an emulator, denoted as $y_{\text{EM}}$, mainly stem from the charge-independence assumption and the SVD truncation.
    The standard deviation $\sigma$ is obtained using Eq.~\eqref{eq:3N_cov}. Experimental results are shown in the last column for comparison.}
    \label{tab:3N_oberservables}
    \renewcommand{\arraystretch}{1.3}
    \begin{ruledtabular}
    \begin{tabular}{l | c | c | c | c}
    & $y_{\text{ref}}$ & $y_{\text{EM}}$ & $\sigma $ & Experiment \\
    \hline
     $E(^3\text{H})$ (MeV)   & $-8.69$  & $-8.48$  & $0.11$ & $-8.48$ \cite{Fiarman1975NPA_TritonBindingExp}   \\
      $fT_{1/2}$ (s)    &  $1228.76$ & $1227.72$ \cite{takayuki_private} & $16.09$ & $1129.6$ \cite{Akulov2005PLB_TritonHalfLife}
    \end{tabular}
    \end{ruledtabular}
\end{table}

\subsection{Sampling the posterior}
\label{sec:PosteriorSampling}

We perform joint inference for $\boldsymbol{\alpha}_{\nn{}}$, $c_D$, and $c_E$ based on NN phase shifts $\mathcal{D}$
and the triton observables $E(^3\text{H})$, $fT_{1/2}$.
For this, we sample from the joint posterior
\begin{align}
&\text{pr}(c_D, c_E, \boldsymbol{\alpha}_{\mathrm{NN}} |\mathcal{D}, E(^3\text{H}), fT_{1/2}) \nonumber \\
&\quad \propto \text{pr}(\mathcal{D}|\boldsymbol{\alpha}_{\mathrm{NN}} )\,\text{pr}(E(^3\text{H}), fT_{1/2} | c_D, c_E, \boldsymbol{\alpha}_{\mathrm{NN}}) \nonumber \\
&\qquad \times \text{pr}(c_D, c_E, \boldsymbol{\alpha}_{\mathrm{NN}}) \,, 
\end{align}
constructed from the NN and 3N likelihoods in Eqs.~\eqref{eq:likelihood_NN_6} and \eqref{eq:likelihood_3N}
and the uniform priors $\text{pr}(c_D, c_E, \boldsymbol{\alpha}_{\mathrm{NN}})$ discussed in the previous sections.
For this, we use the affine Markov chain Monte Carlo (MCMC) sampler \texttt{emcee}~\cite{foreman2013emcee}.

We use $100$ walkers in total that are initialized around the default parameters. Each walker performs 10,000 steps to ensure stationarity, and the first 5,000 are discarded as burn-in.
In total, we then have 500,000 samples from our MCMC sampling.
To reduce this to a manageable amount of samples for nuclear structure studies, we subsample by randomly selecting a single sample from each walker, leaving us with a total of $N=100$ samples. This can be seen as an extreme way of thinning the walkers, another way to reduce autocorrelation.

We find that the full \nn{} posterior depends on the positions of the energies where we evaluate the individual likelihoods.
To investigate the impact of the choice of these positions, we perform two independent runs with different energy grids for Eq.~\eqref{eq:cov_matrix_NN}.
The first one will use the energies
\begin{align}
\mathbf{E}_1 
= \big(&1, 3, 5, 10, 15, 20, 30, 50, 75, 100, \notag \\
       &125, 150, 175, 200 \big) \, \mathrm{MeV} \,,
\label{eq:E1}
\end{align}
emphasizing
low-energy phase shifts. This amounts to a total of $14$ energies at which we evaluate the phase shifts.
In the second run we will use only eight evenly spaced energies 
\begin{equation}
 \mathbf{E}_2 = \big(25, 50, 75, 100, 125, 150, 175, 200 \big) \, \mathrm{MeV} \,.   
 \label{eq:E2}
\end{equation}
Using $\mathbf{E}_2$ should reduce the error due to missing correlations, as we evaluate energies not too close to each other.

Phase-shift uncertainties are orders of magnitude smaller at low energies ($\sim 1$\,MeV) compared to larger energies (>$50$\,MeV), as seen in Table~\ref{table:phase_shift_uncertainties}.
Evaluating the likelihood at such small energies puts strong constraints on the overall possible phase shifts and thereby on the underlying LEC combinations.
We use the second run to investigate the impact of including low energies in the likelihood and also to investigate the influence of broader phase-shift and parameter distributions on the posterior predictive distributions.

\begin{figure*}[t!]
    \centering
    \includegraphics[width=\linewidth]{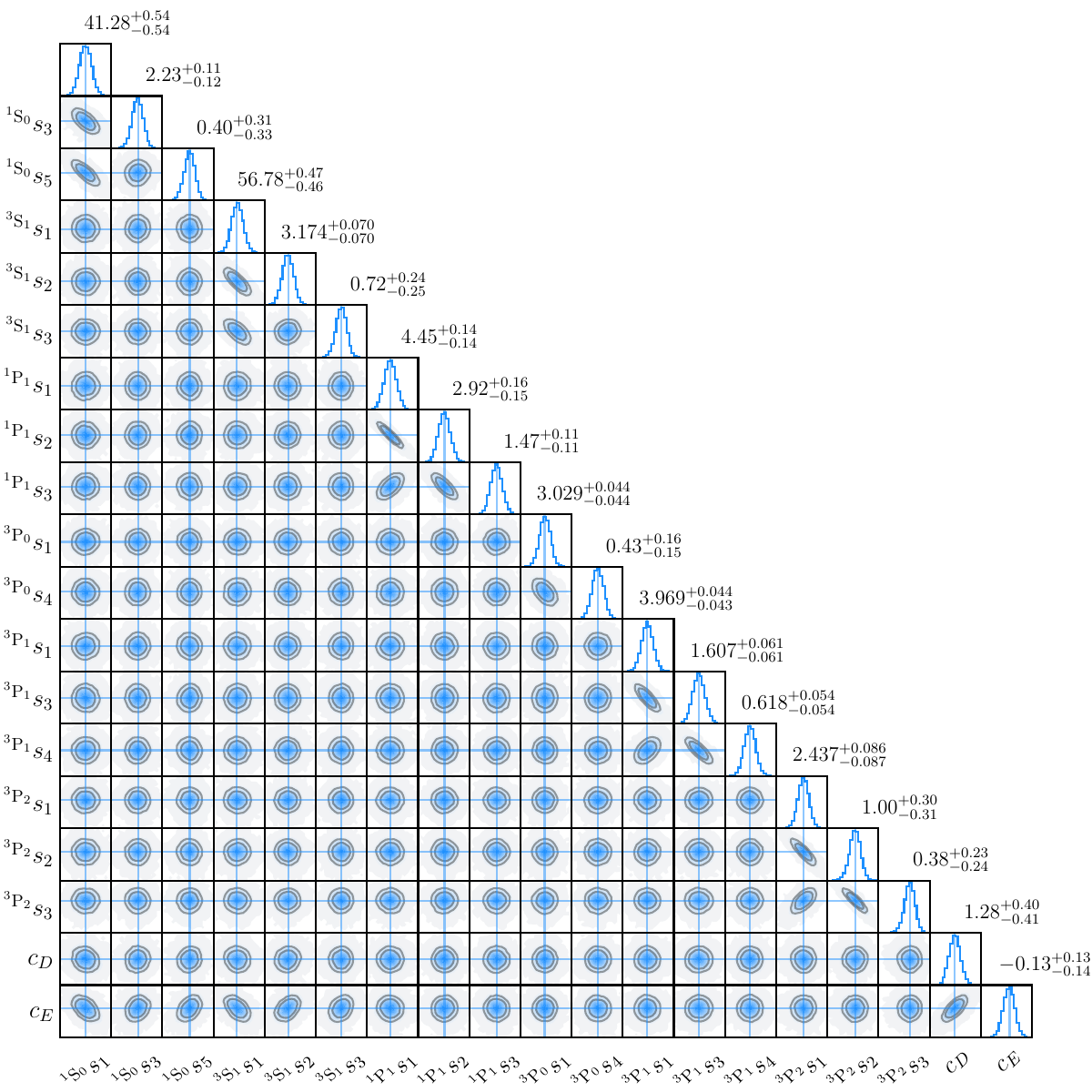}
    \caption{Posterior distributions for the jointly sampled $17$ NN and $2$ \nnn{} parameters for all $5\times10^5$ samples. Distributions were obtained using the $\mathbf{E}_2$ likelihoods. Contours enclose $39\%$ and $87\%$ of the total probability, matching the fractions of a $1\sigma$ and $2\sigma$ contour for a $2$D Gaussian. Values on top denote the median and the marginalized $68\%$ credible intervals. Blue lines represent the default singular values and \nnn{} LECs of the decomposed \magic{} interaction.}
    \label{fig:posterior_parameters}
\end{figure*}

We are left with $N=100$ samples of our two- and three-body parameters $\{\boldsymbol{\tilde{\alpha}}_i\}_{i=1}^N$, where $\boldsymbol{\tilde{\alpha}} = \boldsymbol{ \alpha}_{\mathrm{NN}},c_D, c_E$, according to the posterior distribution pr$(\boldsymbol{\tilde{\alpha}}|\mathcal{D})$. The full posterior distribution is shown in Fig.~\ref{fig:posterior_parameters}.
These samples account for the truncation uncertainty of phase shifts in the S- and P-waves as well as those of the triton binding energy and the comparative half-life.
We observe that the reference singular values and LECs of the \magic{} interaction are reproduced by construction.
Most parameters are uncorrelated as they primarily affect independent partial waves and the 3N observables do not yield strong correlations across partial waves.
We do notice that often the leading singular values in a partial wave are anticorrelated.
In these cases, both terms impact phase shifts in the same energy range, so variations in one singular value must be compensated by the other, leading to the anticorrelation
(see, e.g., $s_1$ and $s_2$ in Fig.~\ref{fig:SVD_phase_shifts}).

The same inference performed using a scale-invariant prior for the NN singular values yields very similar results.
We used a scale-invariant prior of the form
\begin{equation}
    p(s) \propto \begin{cases}
        1/|s| &  10^{-3} < |s| < 100 \,, \\
        1 / 10^{-3} & |s| \leq 10^{-3} \,, \\
        0& \text{otherwise} \,,
    \end{cases}
    \label{eq:prior}
\end{equation}
where we also explored different threshold values.
The piecewise treatment is required as this prior diverges for singular values close to zero. 
The posterior distributions for most singular values are nearly identical.
Only for singular values close to 0 does the prior impact the posterior distribution.
In these cases the posterior peaks closer to 0 and has long asymmetric tails.
This also weakly impacts the inferred distributions for correlated parameters.
Ultimately, the inferred medians and 68\% credible intervals for all parameters are similar for both choices of prior, validating our choice of uniform prior.

\subsection{Posterior predictive distributions}

\begin{figure*}[t!]
    \centering
    \includegraphics[width=\textwidth]{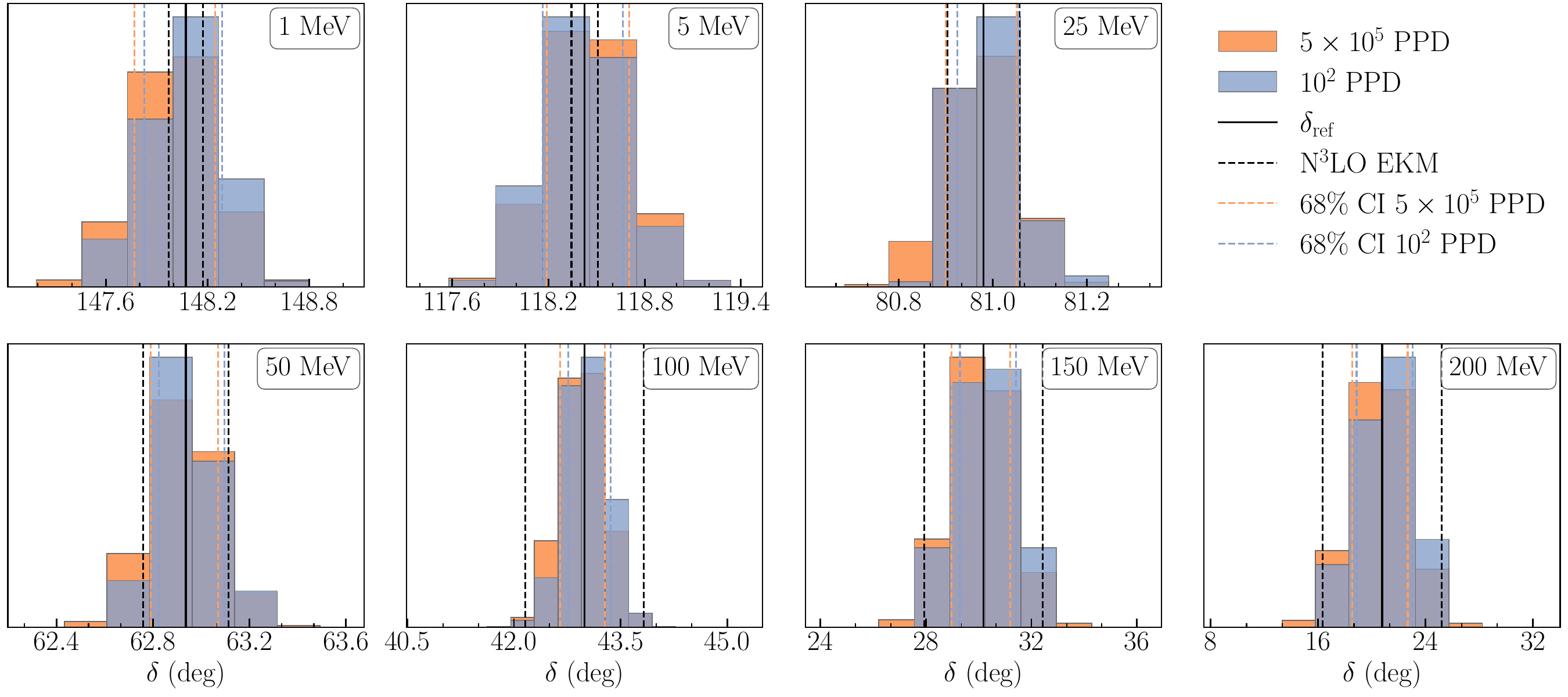}
    \caption{Comparison of phase shifts PPDs in the $^3$S$_1$ partial wave for all $5\times10^5$ samples (orange) and for the resampled $10^2$ samples (blue) at different energies. Results were obtained using the $\mathbf{E}_2$ likelihood.
    The orange and blue dashed lines show the corresponding $68\%$ credible intervals. The solid black line represents the reference phase shifts obtained by using the SVD-decomposed N$^3$LO EM $500$ $\lambda=1.8\,\fminv{}$ interaction at rank five. The black dashed lines represent the EKM uncertainty at N$^3$LO. }
    \label{fig:pre_post_subsampling}
\end{figure*}

In this section, we evaluate our sampling results through posterior predictive distributions (PPDs) of the same observables used for the Bayesian inference of the parameters $\boldsymbol{\tilde{\alpha}}$. This process is part of model checking.
If the likelihood correctly encodes the data and model uncertainties and the priors are uninformative, the posterior distributions should reflect those uncertainties.
We consider both runs using the energy grids $\mathbf{E}_1$ and $\mathbf{E}_2$
and separately analyze the NN and $3$N components of our framework.

\begin{figure*}[t!]
    \centering
    \includegraphics[width=\textwidth]{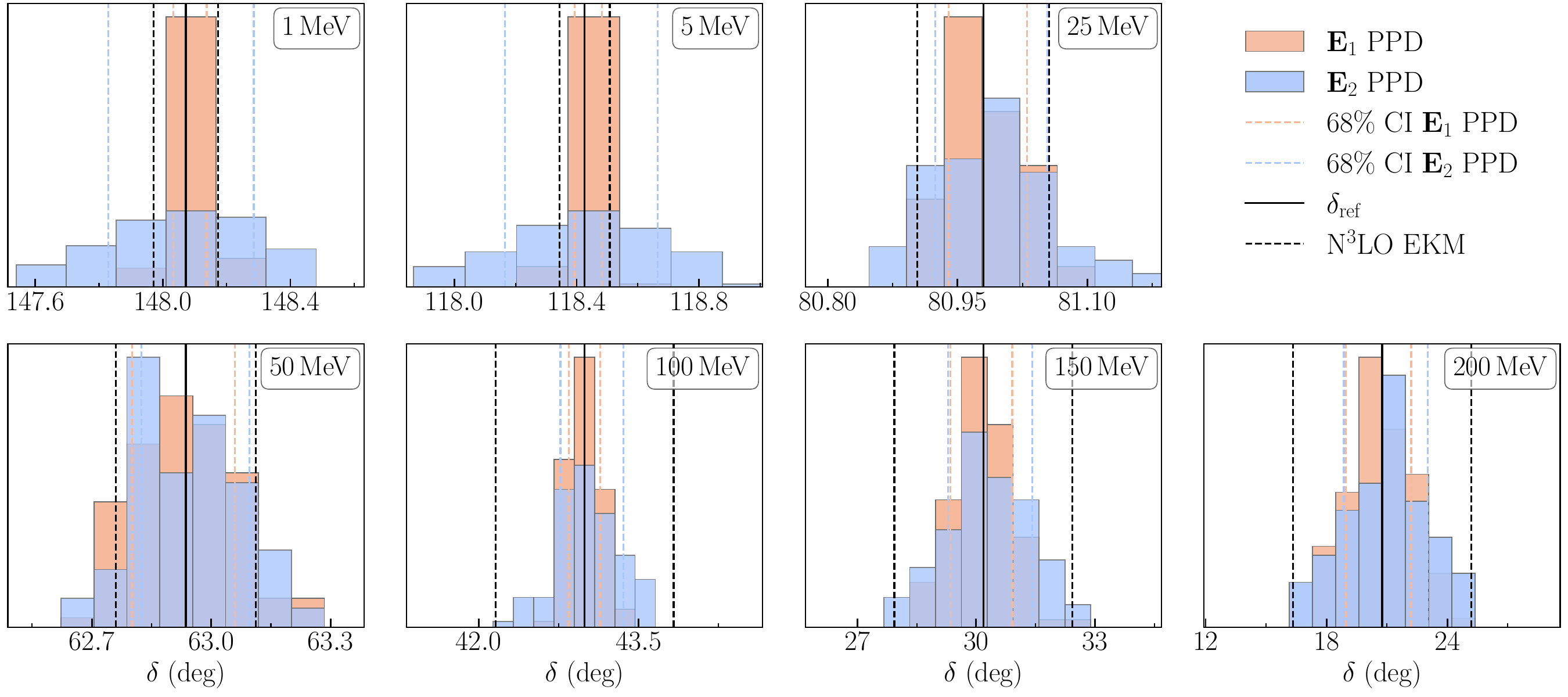}
    \caption{Comparison of phase shifts PPDs in the $^3$S$_1$ partial wave using the $\mathbf{E}_1$ likelihood (orange) and the $\mathbf{E}_2$ likelihood (blue) at different energies. Shown are the $100$ subsamples. The orange and blue dashed lines show the corresponding $68\%$ credible intervals. The solid black line represents the reference phase shifts obtained by using the EM $500$ $\lambda=1.8\,\fminv{}$ interaction at N$^3$LO in SVD decomposition up to rank five. The black dashed lines represent the EKM uncertainty at N$^3$LO. }
    \label{fig:3s1_ppd_phas_shifts}
\end{figure*}

\begin{figure*}[t!]
    \centering
    \includegraphics[width=\textwidth]{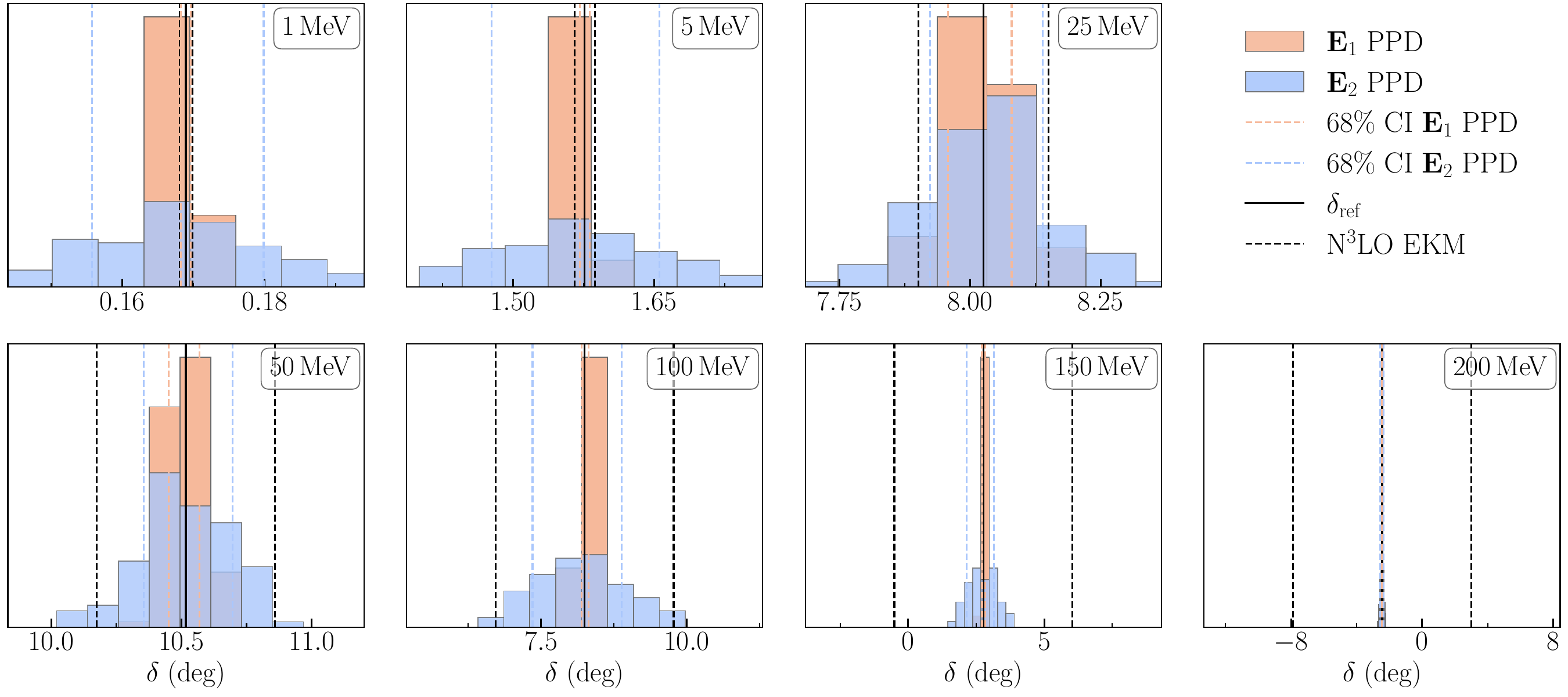}
    \caption{Same as Fig.~\ref{fig:3s1_ppd_phas_shifts} for the $^3$P$_0$ partial wave. }
    \label{fig:3p0_ppd_phas_shifts}
\end{figure*}

\begin{figure*}[t!]
    \centering
    \includegraphics[width=\textwidth]{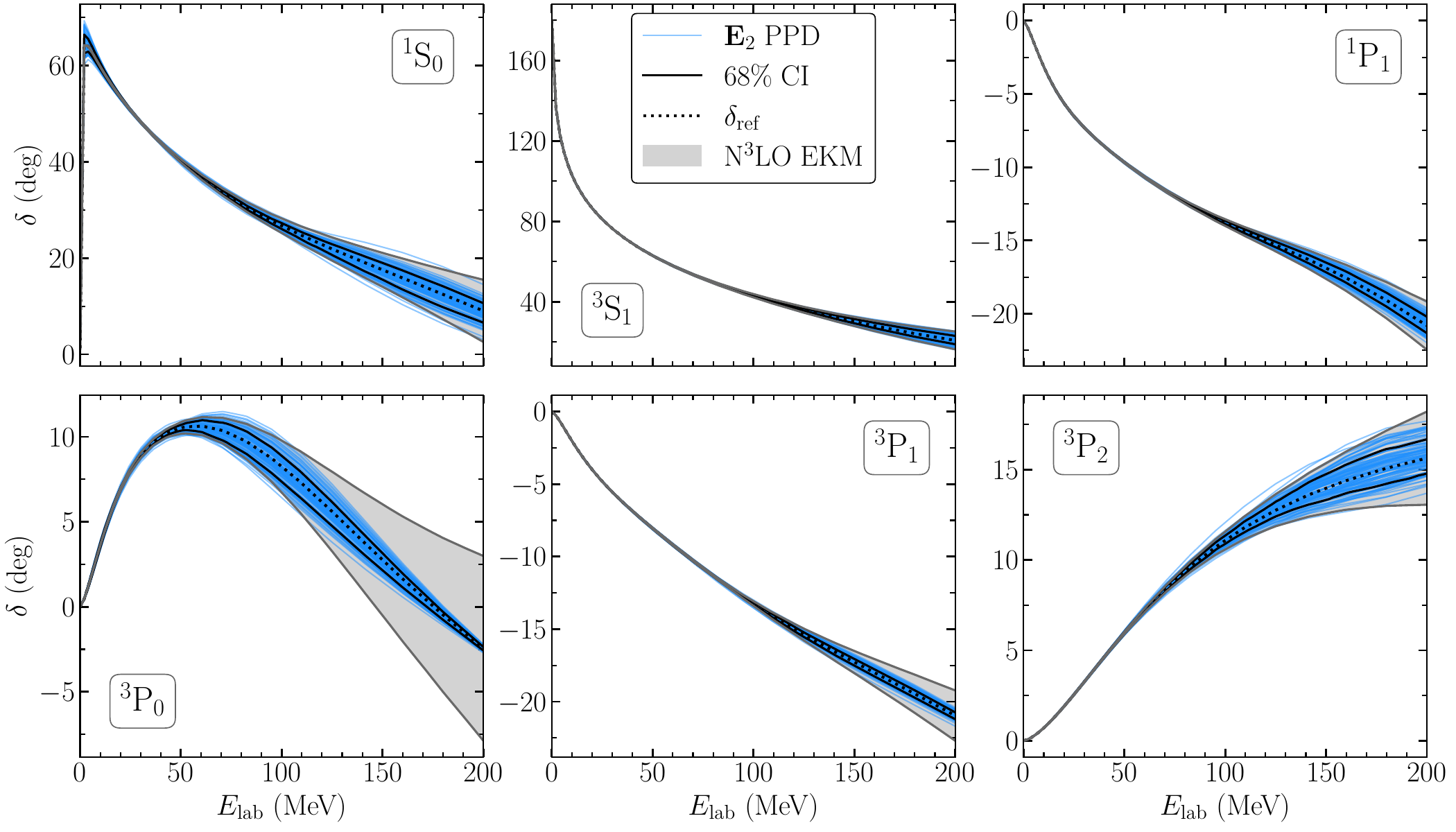}
        \caption{Posterior phase shift distributions for the $100$ subsamples in S- and P-waves based on the $\mathbf{E}_2$ likelihood. The dotted black line represents the reference phase shifts obtained by using the SVD-decomposed N$^3$LO EM $500$ $\lambda=1.8\,\fminv{}$ interaction at rank five. The gray band represents the EKM uncertainty at N$^3$LO. The blue lines represent the individual samples, the black solid line corresponds to their $68\%$ credible intervals.}
    \label{fig:ppds_phase_shifts}
\end{figure*}

\begin{figure}[t!]
    \centering
    \includegraphics[width=\linewidth]{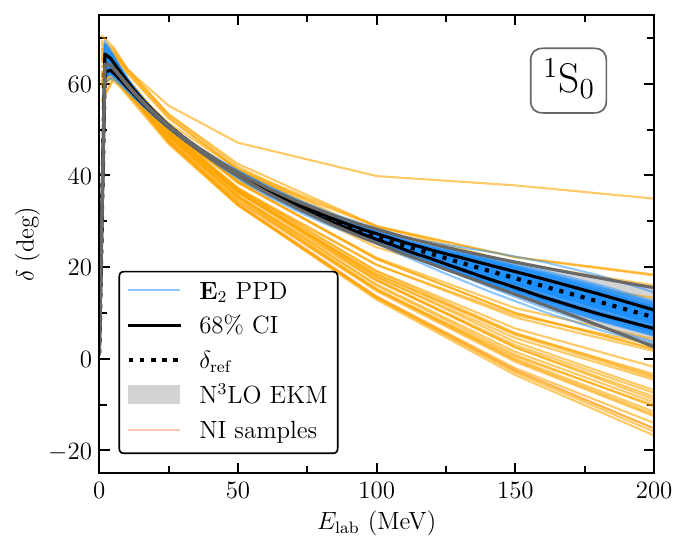}
    \caption{Posterior phase shift distributions for the $100$ subsamples in the $^1$S$_0$ partial wave based on the $\mathbf{E}_2$ likelihood. The dotted black line represents the reference phase shifts obtained by using the SVD-decomposed N$^3$LO EM $500$ $\lambda=1.8\,\fminv{}$ interaction at rank five. The gray band represents the EKM uncertainty at N$^3$LO. The blue lines represent the individual samples, the black solid line corresponds to their $68\%$ credible intervals. Shown in orange are the $34$ nonimplausible (NI) samples obtained from history matching in~\cite{Hu2022NP_Pb208}.
    Note that the NI samples are constructed at N$^2$LO in $\Delta$-full chiral EFT.}
    \label{fig:ppd_1s0}
\end{figure}

\begin{figure*}[t!]
    \centering
        \includegraphics[width=0.4\textwidth]{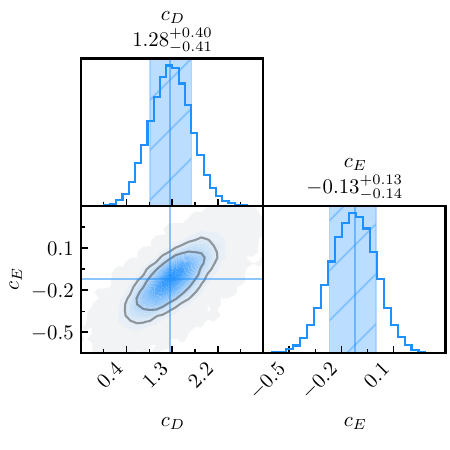}
        \hspace*{1cm}
        \includegraphics[width=0.4\textwidth]{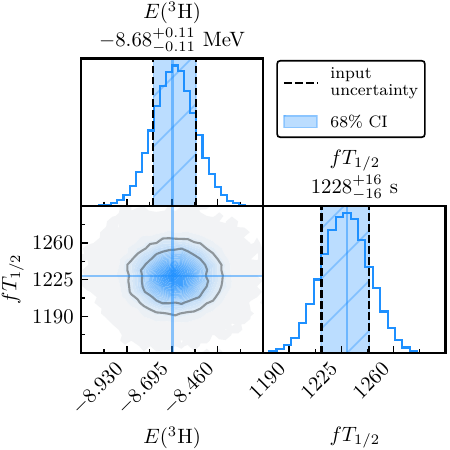} 
    \caption{Posterior distributions of $c_D$ and $c_E$ as well as posterior predictive distributions of $E(^3\mathrm{H})$ and $fT_{1/2}$ for all $5\times10^5$ samples based on the $\mathbf{E}_2$ likelihood. The bottom left panel shows the joint probability distribution of $c_D$, $c_E$ on the left and $fT_{1/2}$, $E(^3\mathrm{H})$ on the right. Contours enclose $39\%$ and $87\%$ of the total probability, matching the fractions of a $1\sigma$ and $2\sigma$ contour for a $2$D Gaussian. The top and right panels show the marginalized distributions of $c_D$ and $c_E$ on the left and  $E(^3\mathrm{H})$ and $fT_{1/2}$ on the right. The blue range shows the $68\%$ CIs of the distributions. The black dashed lines denote the EFT truncation uncertainties used to construct the likelihood.}
    \label{fig:ppd_3N}
\end{figure*}

Using the sampled values for the parameters  $\{\boldsymbol{\tilde{\alpha}}_i\}_{i=1}^N$, we compute the PPDs
\begin{equation}
    \text{PPD} = \{\textbf{y}(\boldsymbol{\tilde{\alpha}}): \boldsymbol{\tilde{\alpha}} \sim\text{pr}(\boldsymbol{\tilde{\alpha}} | \mathcal{D})\} \,,
    \label{eq:ppd}
\end{equation}
to assess the consistency of our parameter distributions.
The posterior $\text{pr}(\boldsymbol{\tilde{\alpha}} | \mathcal{D})$ is the result of the MCMC sampling.
When calculating phase shifts in a given partial wave, we effectively marginalize over $c_D$, $c_E$, and all $\boldsymbol{ \alpha}_{\mathrm{NN}}$ that correspond to different partial waves. The triton observables, on the other hand, depend on both \nn{} and \nnn{} parameters.

In Fig.~\ref{fig:pre_post_subsampling}, we compare PPDs for phase shifts in the $^3$S$_1$ partial wave before and after subsampling for the distributions obtained using the $\mathbf{E}_2$ likelihood.
The individual plots show phase shift distributions at different energies. At each energy we observe the distributions as well as their associated $68\%$ credible intervals (CIs) to be very similar. We observe similar behavior in the other S- and P-waves. This validates our approximation to use only a fraction of the initial samples.

In Fig.~\ref{fig:3s1_ppd_phas_shifts}, we compare phase shift distributions obtained using the $\mathbf{E}_1$ and $\mathbf{E}_2$ likelihood, again in the $^3$S$_1$ partial wave. Our main interest is to see how the posterior distributions compare to the distributions we assumed in the likelihood.
Ideally, the distributions are identical, which we would observe as the EKM uncertainty (black dashed lines) and the $68\%$ credible intervals of the PPDs (orange and blue dashed lines) aligning.

In practice, we observe deviations from these ideal expectations. Phase shift distributions over- and underpredict the EKM uncertainties, depending on the energy.
Additionally, the choice of the likelihood, e.g., the energies at which we evaluate phase shifts in the \nn{} likelihood impacts the posterior distributions of phase shifts.
Considering again Fig.~\ref{fig:SVD_phase_shifts}, we see that each operator affects phase shifts over a range of energies.
We cannot expect to vary phase shifts independently over a range of $200\,\MeV{}$ with only three parameters that are not independent in the energies they impact.
We will see later the reproduction of uncertainties further deteriorates when we only have two instead of three parameters, as is the case for phase shifts in the $^3$P$_0$ partial wave.
We note, however, that simply extending the decomposition to higher ranks and allowing more singular values to vary within a single partial wave does not improve the reproduction of the input uncertainties.
We simply lack the necessary independent degrees of freedom to fully reproduce the EFT uncertainties at all energies,
which is unsurprising given that these are naturally also correlated and thus not fully independent as we conservatively assume.

Coming back to Fig.~\ref{fig:3s1_ppd_phas_shifts}, we observe that at energies below $25\,\MeV{}$, the $\mathbf{E}_2$ results overpredict the EKM uncertainty, reflecting the fact that we put no direct constraints on phase shifts at these energies.
Only correlations with phase shifts at higher energies, which are directly included in the likelihood, constrain these low-energy phase shifts.
In contrast, the $\mathbf{E}_1$ likelihood already actively constrains phase shifts at $1\,\MeV{}$, which yields narrower phase shifts distributions. The distributions using $\mathbf{E}_1$ are narrower than the initial EKM uncertainties.
Going to higher energies, both distributions underpredict the EKM uncertainties, with significantly smaller 68\% CIs starting at $100\,\MeV$. The $\mathbf{E}_2$ distributions are broader than the $\mathbf{E}_1$ distributions. This is again due to correlations: stronger constraints on low-energy phase shifts result in stronger constraints at higher energies.

In Fig.~\ref{fig:3p0_ppd_phas_shifts}, we show the same results in the $^3$P$_0$ partial wave. Note that we only vary two parameters in this partial wave because other operators are either not linearly independent from the two we include or produce no notable variations at energies below $200\,\MeV{}$.
Here we can see the same qualitative behavior as before, only more pronounced. Phase shift uncertainties are strongly overpredicted at low energies for the $\mathbf{E}_2$ likelihood and underpredicted at high energies. Once again the operator structure does not allow for phase shifts to vary independently at different energies. The $\mathbf{E}_1$ distributions describe the EKM uncertainty well at $1\,\MeV{}$ but underpredict it at all higher energies.

We observe similar behavior in all partial waves. At low energies, $\mathbf{E}_2$ distributions overpredict the EFT uncertainties while $\mathbf{E}_1$ distributions  underpredict, and both underpredict EFT uncertainties at energies above $25\,\MeV{}$.
This can be seen in Fig.~\ref{fig:ppds_phase_shifts}, where uncertainties are particularly underpredicted by the PPD at high energies in the $^3$P$_0$ partial wave.
Based on this intuition we treat the $\mathbf{E}_2$ distributions as the more conservative uncertainty, while $\mathbf{E}_1$ distributions serve as the lower end of uncertainties. Note that the EKM uncertainty itself is already a conservative way of estimating uncertainties.

In Fig.~\ref{fig:ppd_1s0}, we show the subsampled PPD for phase shifts in the $^1$S$_0$ partial wave. 
The orange lines correspond to the $34$ nonimplausible samples obtained through history matching from \cite{Hu2022NP_Pb208} 
using $\Delta$-full chiral EFT at N$^2$LO.
Since these samples are obtained by passing a series of implausibility checks,
they do not have a statistical interpretation.
Instead, they must be resampled according to a separate likelihood to provide a PPD.
We compare to the posterior distribution of our $\mathbf{E}_2$ likelihood,
noting that EFT uncertainties at different orders (N$^2$LO for the nonimplausible samples vs.~N$^3$LO in this work) are employed in the construction.
This makes direct comparison challenging, but we gain insight into the effect of going to higher orders in chiral EFT and of using a more stringent likelihood sampling approach as compared to the conservative Pukelsheim's $3\sigma$ nonimplausibility criterion used in~\cite{Hu2022NP_Pb208}.
We conclude that our distribution is significantly more constrained than the nonimplausible samples across all energies considered.
Finally we turn our attention to the reproduction of input EFT uncertainties for our \nnn{} observables.
Figure \ref{fig:ppd_3N} shows the posteriors for $c_D$ and $c_E$ marginalized over all \nn{} parameters, as well as the PPDs for $E(^3{\mathrm{H}})$ and $fT_{1/2}$ for all $5\times10^5$ samples. These distributions shown here were obtained using the $\boldsymbol{E}_2$ likelihood, however the results for $\boldsymbol{E}_1$ are not noticeably different. The distributions for $E(^3{\mathrm{H}})$ and $fT_{1/2}$ are essentially uncorrelated, further justifying the use of these two observables to constrain $c_D$ and $c_E$. The black dashed lines represent one standard deviation of the uncertainty used to construct the \nnn{} likelihood, as defined in Eq.~\eqref{eq:3N_cov}. 
The blue band represents the $68\%$ credible interval of the PPD for the observables. The two intervals are in excellent agreement, indicating that our samples are distributed according to the \nnn{} likelihood we constructed.
\section{Applications to medium-mass nuclei}
\label{sec:MediumMassApplications}

With two different posterior distributions for the parametric uncertainty of the $1.8/2.0$ (EM) interaction at hand, we investigate their impact on nuclear structure calculations. We are in particular interested in quantifying the uncertainties in past predictions using the 1.8/2.0 (EM) Hamiltonian and comparing with other state-of-the-art approaches to quantifying uncertainties in nuclear structure calculations~\cite{Hagen2016NP_Ca48Skin, Huther2020N3LO, Hu2022NP_Pb208, Kondo2023N_O28, Maris2022PRC_LENPICSMSNuclei}. 
Other approaches assume zero correlation between the parametric uncertainty captured in the distribution of LECs and the EFT truncation uncertainty of nuclear structure observables.
This results in an additive model of uncertainties
\begin{equation}
    y = y_k + \delta_{\mathrm{model}} +  \delta_{\mathrm{method}} \,,
\end{equation}
where $y_k$ contains parametric uncertainties, $\delta_{\mathrm{model}}$ is the EFT truncation uncertainty, and $\delta_{\mathrm{method}}$ the uncertainty due to other effects like many-body truncations.
In our approach,
the input LECs are fixed and only the singular values $s_i$ are uncertain.
Hence our ``parametric uncertainty'' is not driven by fitting errors but is constructed from truncation estimates in the NN and 3N sectors. In this setup truncation errors are already captured by the posterior distribution of singular values resulting from our Bayesian inference.
Assuming perfect correlation between truncation uncertainties of these NN and 3N systems and medium-mass nuclei observables, we would be double counting uncertainties if we included an additional $\delta_{\mathrm{model}}$ term for the target observable of interest, leading to too conservative uncertainties for nuclear structure predictions.

In this work, we focus on the parametric uncertainty, which we expect to capture a large part of the EFT truncation uncertainty in nuclear structure observables, and do not consider an additional explicit EFT truncation uncertainty $\delta_\mathrm{model}$.
Note that we also do not consider a $\delta_{\mathrm{method}}$ in this work.
Under these assumptions our uncertainty model reduces to just the posterior predictive distribution from Eq.~\eqref{eq:ppd}, i.e., the propagation of truncation-induced parametric uncertainties to target observables.
Many bulk observables in nuclei are correlated, and so are the systematic theory uncertainties when computing them~\cite{ Hagen2016NP_Ca48Skin, 
Neufcourt:2018syo, Hu2022NP_Pb208, Maris2022PRC_LENPICSMSNuclei, Heinz2025PRC_ImprovedCalcium, Heinz2024inprep_MuToEResponses, Miyagi2025arxiv}. 
Developing a quantitative understanding how to handle correlations between the parametric and EFT truncation uncertainty is an important avenue for future work that is necessary for complete and accurate uncertainty quantification.

We compute the structure of oxygen and calcium isotopes using the in-medium similarity renormalization group (IMSRG)~\cite{Tsukiyama2011PRL_IMSRG, Hergert2016PR_IMSRG}, a well established quantum many-body method, truncated at the IMSRG(2) level. All calculations are performed with the \textsc{imsrg++} code~\cite{Stroberg2024_IMSRGGit}.
\nn{} and \nnn{} matrix elements are expanded in a spherical harmonic-oscillator basis with frequency $\hbar \omega = 16\,$MeV and employing a truncation on the single-particle basis based on $e=2n+l \leq e_{\mathrm{max}}=14$.
For 3N matrix elements we employ an additional truncation on the three-body basis, including states $\ket{123}$ with $e_1 + e_2 + e_3 \leq E^{(3)}_{\mathrm{max}} = 24$~\cite{Miyagi2022PRC_NO2B}.
We use an optimized basis consisting of Hartree-Fock and perturbatively improved natural orbital states following the construction of Ref.~\cite{Heinz2025PRC_ImprovedCalcium}.
Finally, we apply a truncation of $e_{\mathrm{max}}=10$ to the natural orbitals.

We limit our investigations to oxygen and calcium isotopes in this work although the 100 samples can be applied to any nuclear structure calculation that is based on the \magic{} interaction. 
We investigate the impact of our different assumptions by calculating $^{24}$O, $^{28}$O, and $^{48}$Ca observables using different variations of the \magic{} interaction.
First, we use the unchanged \magic{} interaction. Then we use the same interaction but with SVD truncated S- and P-waves at $R_{\mathrm{SVD}} = 5$. We also investigate the effect of the charge independence assumption in the S- and P-waves on our observables by applying Eq.~\eqref{eq:charge_independence} to the S- and P-wave interactions.
In a third variation we combine the two effects.

\begin{table*}[t!]
\renewcommand{\arraystretch}{1.3}
\centering
\caption{Computed observables for variations of the \magic{} interaction: ``Unchanged'' using the original \magic{} Hamiltonian; ``SVD'' after applying the rank-five SVD truncation in the S- and P-waves;  ``CII'' using the charge-independent interaction approximation for the S- and P-waves; and ``SVD \& CII'' with the two previous approximations combined. See main text for details. $\Delta$ indicates the difference to the result using the original \magic{} Hamiltonian.}
\label{tab:observables}
\begin{ruledtabular}
\setlength\tabcolsep{1em}
\begin{tabular}{l|l|rr|rr|rr}
Observable & Configuration & \multicolumn{2}{c|}{$^{24}$O} & \multicolumn{2}{c|}{$^{28}$O} & \multicolumn{2}{c}{$^{48}$Ca} \\
&  & Value & $\Delta$ & Value & $\Delta$ & Value & $\Delta$ \\ 
\hline
$E$ (MeV) 
& Unchanged      &   $-164.01$ &    ---    &   $-162.45$ &    ---    &   $-415.77$ &    ---    \\ 
& SVD     &   $-164.12$ &   $-0.11$   &   $-162.59$ &   $-0.14$   &   $-416.12$ &   $-0.35$   \\ 
& CII      &   $-168.59$ &   $-4.58$   &   $-167.39$ &   $-4.94$   &   $-426.39$ &  $-10.62$   \\ 
& SVD \& CII    &   $-168.70$ &   $-4.69$   &   $-167.53$ &   $-5.08$   &   $-426.75$ &  $-10.98$   \\ 
\hline
$R_{\mathrm{skin}}$ (fm) 
& Unchanged      &    0.4764 &    ---    &    0.6713 &    ---    &    0.1439 &    ---    \\ 
& SVD      &    0.4762 &  $-0.0002$  &    0.6709 &  $-0.0004$  &    0.1439 &   0.0000  \\ 
& CII      &    0.4730 &  $-0.0034$  &    0.6667 &  $-0.0046$  &    0.1444 &   0.0005  \\ 
& SVD \& CII    &    0.4728 &  $-0.0035$  &    0.6663 &  $-0.0050$  &    0.1444 &   0.0005  \\ 
\hline
$R_{\mathrm{ch}}$ (fm) 
& Unchanged      &    2.611  &    ---    &    2.765  &    ---    &    3.290  &    ---    \\ 
& SVD      &    2.611  &  $0.000$   &    2.764  &  $0.000$   &    3.290  &   0.000   \\ 
& CII      &    2.599  &  $-0.012$   &    2.753  &  $-0.012$   &    3.276  &  $-0.013$   \\ 
& SVD \& CII    &    2.599  &  $-0.012$   &    2.752  &  $-0.012$   &    3.277  &  $-0.013$   
\end{tabular}
\end{ruledtabular}
\end{table*}

Table~\ref{tab:observables} gives the ground-state energy, charge radius, and neutron skin thickness of $^{24}$O, $^{28}$O, and $^{48}$Ca for the different modifications of the $1.8/2.0$ (EM) interaction, together with the difference to the predictions using the unchanged $1.8/2.0$ (EM) interaction.
We notice that uncertainties caused by the SVD truncation are always below $1\%$. The charge-independence assumption causes deviations of up to $3\%$ for energies. The deviations for charge radii and the neutron skins are still below $1\%$ for all nuclei considered.
Since we are more concerned with uncertainty estimates than reproducing the exact values for observables, this accuracy is sufficient.
The uncertainty due to the SVD truncation can be easily improved in the future by simply adding the higher-order parts of the SVD that are discarded at the moment back to the interaction. 
For the charge-independence assumption, we avoid the explicit treatment of individual isospin channels and instead shift the median values of the distributions to the predictions using the unchanged \magic{} interaction. 
\begin{figure}[t!]
    \centering
    \includegraphics[width=0.875\linewidth]{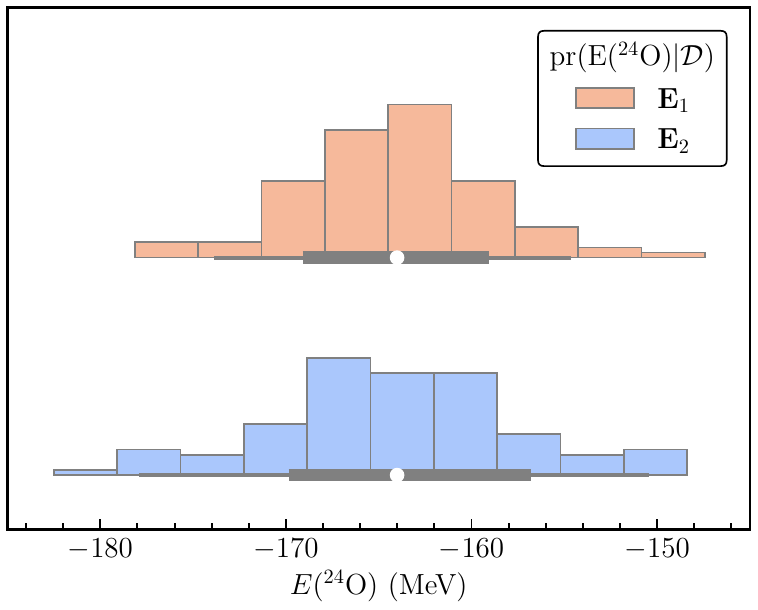}
    \caption{PPDs for the ground-state energy of \elem{O}{24} from the $\mathbf{E}_1$ and $\mathbf{E}_2$ likelihoods. The white dot shows the median while the thick and thin gray bars represent the $68\%$ and $95\%$ credible intervals respectively. Note that the medians were shifted to reproduce the \magic{} interaction predictions.}
    \label{fig:histogram_O24}
\end{figure}

In Fig.~\ref{fig:histogram_O24}, we show the PPDs for the ground-state energy of $^{24}$O. The distribution corresponding to the $\mathbf{E}_1$ likelihood is narrower than the $\mathbf{E}_2$ distribution, as expected.
The resulting $68\%$ credible intervals around the shifted median are $E(^{24}\mathrm{O})_{\mathbf{E}_1}  = -164.01^{+4.98}_{-5.09}\,\MeV $ and $E(^{24}\mathrm{O})_{\mathbf{E}_2}  = -164.01^{+7.22}_{-5.85} \,\MeV $.
We calculate the ratio of the widths of the two distributions for any interval $[x^{\mathrm{low}}, x^{\mathrm{high}}]$, as $r = (x^{\mathrm{high}}_{\mathbf{E}_1} - x^{\mathrm{low}}_{\mathbf{E}_1}) / (x^{\mathrm{high}}_{\mathbf{E}_2} - x^{\mathrm{low}}_{\mathbf{E}_2})$. Here, we get $r_{68\%} = 0.77$ and $r_{95\%} = 0.70$. The two distributions only differ in the construction of their \nn{} likelihoods.
This shows us that the \nn{} uncertainty significantly contributes to the uncertainty of the ground-state energy of $^{24}$O, even though the uncertainties of the N$^3$LO \nn{} interaction go as $\left(Q/\Lambda_b\right)^5$ and the N$^2$LO \nnn{} uncertainties as $\left(Q/\Lambda_b\right)^4$.   
We compare our distributions to the results from~\cite{Hoppe2019PRC_ChiralMedMass}. Their uncertainties were obtained through a direct estimation of the EKM uncertainty of the binding energy of $^{24}$O by assuming $Q = m_{\pi}$ and $\Lambda_b=500\,\MeV$. Hence, these are purely chiral EFT truncation uncertainties.
At N$^3$LO they receive $E(^{24}\mathrm{O}) = -118.5^{+6.24}_{-6.24}\,\MeV$. Note that the median value is significantly different than ours because of the different interaction used. The uncertainties are in a comparable range to ours, but it is important to remember that we consider a mixed order of N$^2$LO and N$^3$LO uncertainties, while~\cite{Hoppe2019PRC_ChiralMedMass} provides an NN+3N N$^3$LO uncertainty estimate.

\begin{table}[t!]
\centering
\renewcommand{\arraystretch}{1.5}
\caption{Uncertainty quantified predictions for $^{24}$O, $^{28}$O, and $^{48}$Ca observables. The results are summarized by the medians and the $68\%$ credible intervals. In this table, results are not shifted to the median of the unchanged \magic{} Hamiltonian.}
\label{tab:distributions}
\begin{ruledtabular}

\begin{tabular}{l|r|r|r}
Nucleus & \multicolumn{1}{c|}{Observable}                        & \multicolumn{1}{c|}{$\mathbf{E}_{1}$}                         & \multicolumn{1}{c}{$\mathbf{E}_{2}$}                         \\
\hline
$^{24}$O    & $E$ (MeV)         & $-167.83^{+4.98}_{-5.09}$     & $-166.87^{+7.22}_{-5.85}$     \\
\hline
$^{28}$O   & $E$ (MeV)         & $-166.84^{+4.43}_{-4.71}$     & $-165.76^{+5.89}_{-5.65}$     \\
\hline
$^{28}\mathrm{O}-{}^{24}$O & $\Delta E$ (MeV) & $\,1.06^{+0.45}_{-0.41}$     & $\,1.13^{+0.81}_{-0.97}$     \\
\hline
$^{48}$Ca   & $E$ (MeV)         & $-424.45^{+12.07}_{-13.61}$   & $-421.61^{+19.43}_{-16.78}$   \\
                 & $R_{\mathrm{skin}}$ (fm)                     & $0.1455^{+0.0018}_{-0.0017}$     & $0.1449^{+0.0027}_{-0.0016}$     \\
                 & $R_{\mathrm{ch}}$ (fm)            & $3.288^{+0.032}_{-0.038}$        & $3.288^{+0.058}_{-0.048}$ 
\end{tabular}    
\end{ruledtabular}
\end{table}

\begin{figure}[t!]
    \centering
    \includegraphics[width=0.875\linewidth]{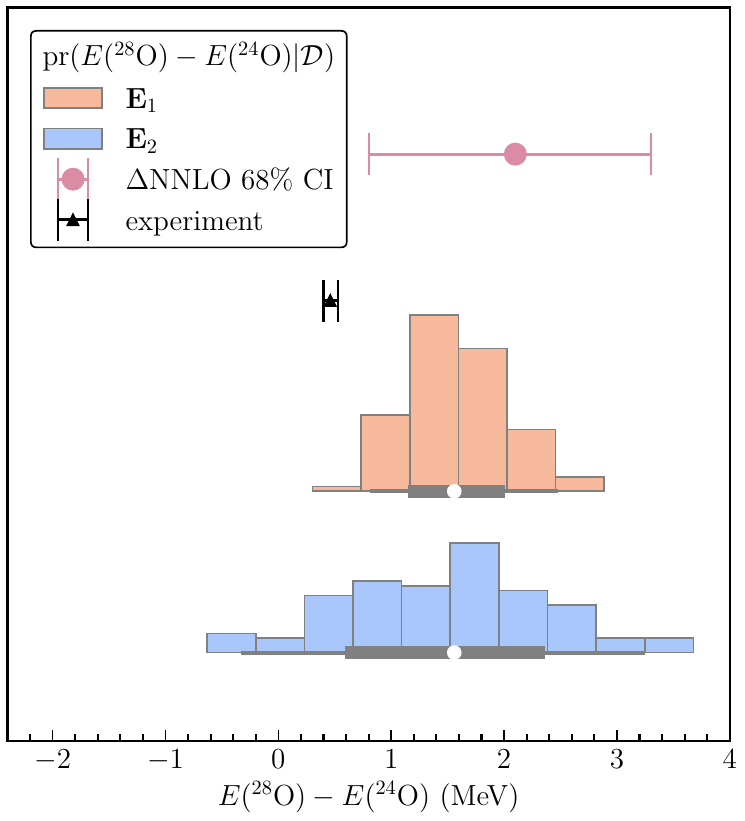}
    \caption{PPDs for the difference between the ground-state energies of \elem{O}{28} and \elem{O}{24} from the $\mathbf{E}_1$ and $\mathbf{E}_2$ likelihoods. The white dot shows the median while the thick and thin gray bars represent the $68\%$ and $95\%$ credible intervals respectively. Note that the medians were shifted to reproduce the \magic{} interaction predictions. The black triangle is experiment~\cite{Kondo2023N_O28}, and the pink dot and error bar represent the median and $68\%$ credible interval from~\cite{Kondo2023N_O28}.}
    \label{fig:histogram_O28-O24}
\end{figure}

We also calculate the distribution of $E(^{28}\mathrm{O})$; see Table\,\ref{tab:distributions}.
By subtracting the ground-state energies of the two oxygen isotopes, we get the 4-neutron separation energy. We find that $^{28}\mathrm{O}$ is a slightly unbound system,
but we note that our calculations do not include continuum effects in the description of either \elem{O}{24} or \elem{O}{28}.
Figure~\ref{fig:histogram_O28-O24} shows the difference in ground-state energy between $^{28}$O and $^{24}$O. 
Here, the difference between the $\mathbf{E}_1$ and $\mathbf{E}_2$ distribution is even more significant with
$\Delta E(^{28, 24}\mathrm{O})_{\mathbf{E}_1} = 1.56^{+0.45}_{-0.41}\,\MeV$ and $\Delta E(^{28, 24}\mathrm{O})_{\mathbf{E}_2}= 1.56^{+0.81}_{-0.97}\,\MeV$, where we shifted the median values to the \magic{} prediction.
The $\mathbf{E}_2$ distribution is twice as wide as the $\mathbf{E}_1$ distribution in terms of the $68\%$ and $95\%$ intervals, with $r_{68\%} = 0.49$, $r_{95\%} = 0.47$.
The experimental value $\Delta E(^{28,24}\mathrm{O}) = 0.46^{+0.05}_{-0.04}(\mathrm{stat})\pm0.02(\mathrm{syst})\,\MeV{}$ falls slightly outside of the $68\%$ credible interval of our $\mathbf{E}_2$ distribution.
For the $\mathbf{E}_1$ distribution, the experimental value is more than $2\sigma$ away $\Delta E(^{28, 24}\mathrm{O})_{\mathbf{E}_1} = 1.56^{+0.45}_{-0.41}\,\MeV$. Again, continuum effects are not taken into account. Note that the deviation due to the charge-independence assumption for $\Delta E(^{28,24}\mathrm{O})$ is significantly larger than for the individual energies, because energy differences are smaller so that isospin symmetry breaking corrections are a larger effect for differential observables.

We additionally compare to the prediction from~\cite{Kondo2023N_O28} using coupled-cluster computations, $\Delta E(^{28, 24}\mathrm{O})= 2.1^{+1.2}_{-1.3}\,\MeV$.
In that work, iterative history matching is performed with observables up to $^{24}$O (NN phase shifts, bulk properties of H and He isotopes, and selected observables in oxygen isotopes)
including experimental, EFT, many-body method, and emulator uncertainties.
After history matching, a likelihood based on nuclei with $A=2$--24 is used to perform a Bayesian inference for the LECs of $\Delta$-full chiral EFT at N$^2$LO
with a Bayesian update to include also information on the experimental separation energy of $^{25}$O relative to $^{24}$O.
The resulting samples are used to compute the PPD for $\Delta E(^{28, 24}\mathrm{O})$.
In comparison, our approach only accounts for EFT uncertainties.

\begin{figure}[!t]
\centering
\includegraphics[width=0.875\columnwidth]{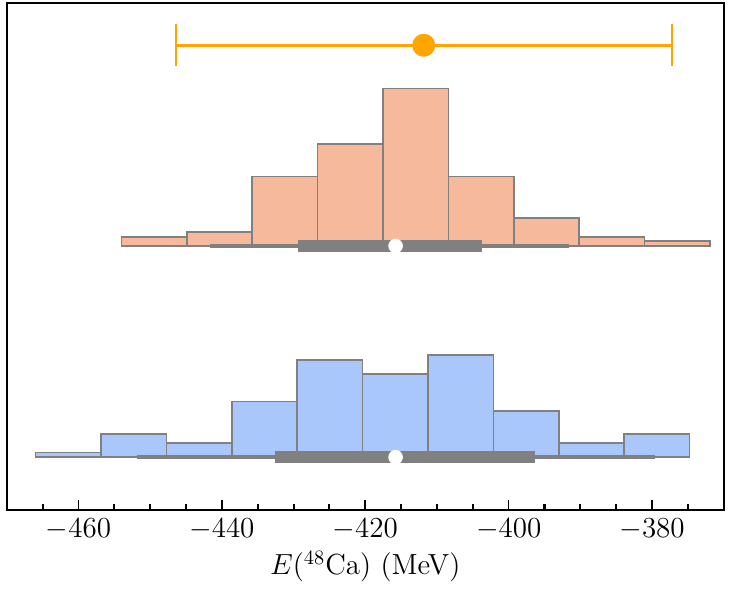}%
\hfill
\includegraphics[width=0.875\columnwidth]{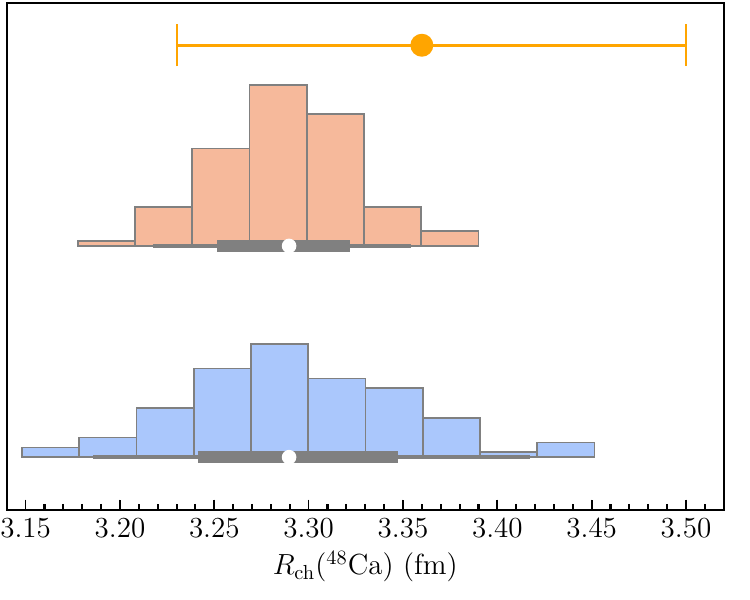}%
\hfill
\includegraphics[width=0.875\columnwidth]{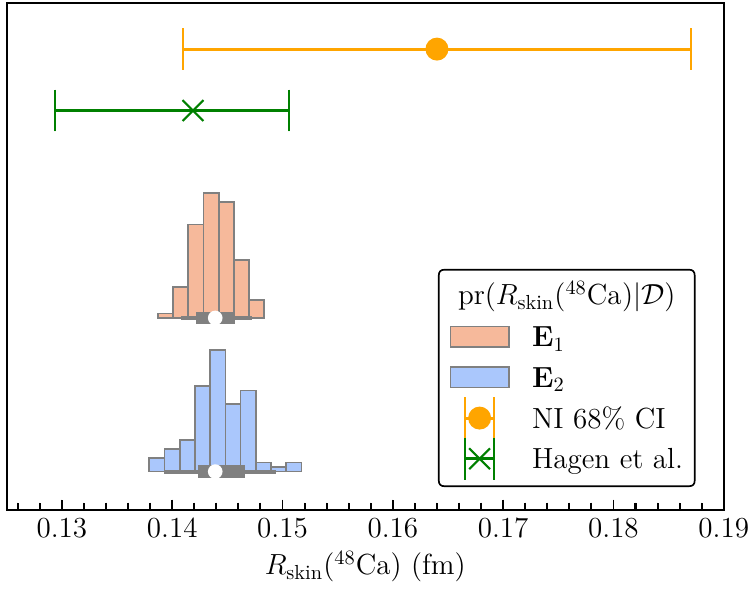}%
\caption{PPDs for the ground-state energy (top panel), charge radius (middle panel), and neutron skin (bottom panel) of \elem{Ca}{48} using the $\mathbf{E}_1$ and $\mathbf{E}_2$ results. The white dot shows the median while the thick and thin gray bars represent the $68\%$ and $95\%$ credible intervals respectively. Note that the medians were shifted to reproduce the \magic{} interaction results. The orange dot and error bar represent the median and $68\%$ credible interval of the $34$ importance resampled nonimplausible (NI) results from~\cite{Hu2022NP_Pb208}. The green band shows the range of \textit{ab initio} predictions using six chiral interactions from~\cite{Hagen2016NP_Ca48Skin}.}
\label{fig:ca48_distributions}
\end{figure}

For $^{48}$Ca, we consider the charge radius and the neutron skin thickness in addition to the ground-state energy. The charge radius is calculated as
\begin{equation}
    R_{\mathrm{ch}}^2 = \braket{R_{\mathrm{p}}^2} + \braket{r_{\mathrm{so}}^2} + r_{\mathrm{p}}^2 + \frac{N}{Z}r_{\mathrm{n}}^2 + \frac{3}{4M^2} \,,
\end{equation}
with the squared point-proton radius $R_{\mathrm{p}}^2$, the spin-orbit correction $r_{\mathrm{so}}^2$~\cite{Ong2010PRC_SpinOrbit,Heinz2025PRC_ImprovedCalcium}, the Darwin-Foldy correction $3/(4M^2) = 0.033\,\fm{}^2$~\cite{Friar1997PRA_DarwinFoldy}, the squared proton charge radius $r_\mathrm{p}^2 = 0.707\,\fm{}^2$, and the neutron charge radius $r_\mathrm{n}^2 = -0.115\,\fm{}^2$~\cite{Workman2022PTEP_PDG2022}.
We compute the neutron skin 
\begin{equation}
    R_{\mathrm{skin}} = \braket{R_{\mathrm{n}}^2}^{1/2}-\braket{R_{\mathrm{p}}^2}^{1/2} \,,
\end{equation}
with the root-mean-square point-neutron and point-proton radii.
The values $\braket{r_{\mathrm{so}}^2}$, $\braket{R_{\mathrm{p}}^2}$, and $\braket{R_{\mathrm{n}}^2}$ are obtained from the IMSRG by transforming the corresponding operators with the same unitary transformation as the Hamiltonian using the Magnus expansion~\cite{Morris2015PRC_Magnus}.

In Fig.~\ref{fig:ca48_distributions}, we provide the posterior distributions for the ground-state energy, the charge radius, and the neutron skin thickness of $^{48}$Ca.
Comparing the distributions from the $\mathbf{E}_1$ and $\mathbf{E}_2$ likelihoods, the $\mathbf{E}_1$ distributions have roughly $70\%$ of the width of the $\mathbf{E}_2$ distributions for all three observables.

We compare with past results from \textit{ab initio} coupled-cluster computations: the $68\%$ credible intervals from~\cite{Hu2022NP_Pb208} based on importance resampled nonimplausible Hamiltonians in orange,
and the range of predictions from six chiral interactions from~\cite{Hagen2016NP_Ca48Skin} in green.
These results from~\cite{Hu2022NP_Pb208} are obtained using history matching on NN phase shifts and bulk properties of nuclei up to $^{16}$O,
which yielded 34 nonimplausible samples,
and importance resampling using a calibration likelihood based on the ground-state energy, point-proton radius, and $2^+$ excitation energy of $^{48}$Ca.
The likelihood employed accounts for experimental, EFT, and many-body method uncertainties.
The final PPDs are constructed from the parametric posterior obtained from importance resampling and independent EFT and many-body uncertainties of the respective observable.
We note that in Fig.~\ref{fig:ca48_distributions} the ground-state energy and charge radius of $^{48}$Ca are calibration observables,
while the neutron skin is a prediction. This causes EFT and many-body uncertainties of the respective observable to enter the final PPDs twice in the case of calibration observables.
For comparison, we do not further modify our parametric PPDs to account for additional uncertainties.

For all three observables our distributions overlap with the nonimplausible distributions, and for the neutron skin with those from the different chiral interactions.
The 68\% credible intervals for the predictions from history matching and importance resampling are significantly wider than those of the $\mathbf{E}_1$ and $\mathbf{E}_2$ PPDs.
This stems from multiple sources.
The nonimplausible samples are constructed at N$^2$LO in chiral EFT,
with a larger EFT uncertainty than the N$^3$LO uncertainties used in this work.
The PPDs from the nonimplausible samples also account for many-body method uncertainties, which we do not include.
The uncertainties of the IMSRG(2) for $^{48}$Ca were explored in~\cite{Heinz2025PRC_ImprovedCalcium} and found to be around $3\%$, $1.5\%$, and $7.5\%$ for ground-state correlation energies, charge radii, and neutron skins.
Accounting for these many-body uncertainties would not significantly change the widths of the distributions for the ground-state energy and charge radius,
but would significantly broaden the distributions for the neutron skin.

The resulting credible intervals with median values not shifted to the unchanged \magic{} results for the different oxygen and calcium observables are given in Table~\ref{tab:distributions}.

\section{Summary and conclusion}
\label{sec:Conclusion}

We have developed a framework to quantify EFT truncation uncertainties for low-resolution interactions. We obtained a linear operator structure for SRG-evolved \nn{} interactions through singular value decompositions.
We used the sensitivity of low-energy phase shifts to identify relevant low-energy operators.
With a parametric description of the interaction at hand, we performed Bayesian inference for the underlying singular values. We constructed the likelihood for the inference from predictions for NN scattering phase shifts in S- and P-waves and for the triton energy and comparative half-life.
For NN phase shifts, we considered two likelihoods, both using phase shifts at energies up to $200\,\MeV{}$.
Phase shifts at different energies are treated as uncorrelated quantities, a simplification that can be improved in future work through the construction of correlated EFT truncation models using Gaussian processes.
One likelihood, $\mathbf{E}_2$, is more conservative than the other $\mathbf{E}_1$, including less information about low-energy phase shifts in the construction.
Accounting for correlations will reduce the effective amount of data, especially in the low-energy region relevant to the $\mathbf{E}_1$ likelihood.
We expect this to ultimately increase the uncertainties in the final inferred posterior distributions when properly accounted for.
Comparing the results using the two likelihoods, $\mathbf{E}_1$ vs.\ $\mathbf{E}_2$, allows us to probe our sensitivity to these missing correlations.

Based on the combined \nn{} and \nnn{} likelihoods, we performed Bayesian inference for the singular values in our \nn{} potentials and for $c_D$ and $c_E$ in our \nnn{} potentials.
We performed model checking for the inferred parameter distributions.
We found that the input \nnn{} likelihood is accurately reproduced by our posterior Hamiltonian distributions.
On the other hand, we found that for the \nn{} phase shifts they did not exactly reproduce the likelihood as expected for unconstrained uniform priors. 
At laboratory energies above $100\,\MeV{}$ both posteriors underpredict the input EFT uncertainties.
We concluded that the limited amount of independent operators in each partial wave and untreated correlations in phase shifts are responsible for these shortcomings.

Finally, we propagated the Hamiltonian uncertainties to medium-mass nuclei through IMSRG calculations using Hamiltonians sampled from our posterior distribution.
We calculated PPDs for the ground-state energies of $^{24}$O, $^{28}$O, and the difference $\Delta E(^{28, 24}\mathrm{O})$.
We found that our more conservative distribution based on the $\mathbf{E}_2$ likelihood is compatible with experiment for $\Delta E(^{28, 24}\mathrm{O})$. 
We also calculated PPDs for the energy, charge radius, and neutron skin thickness of $^{48}$Ca. We compared our distributions to existing estimates of uncertainties for these observables. Our uncertainties were significantly smaller than the more conservative history matching results.

Our work focuses on capturing EFT truncation uncertainties in the distribution of underlying parameters in our Hamiltonian, commonly called parametric uncertainty.
Through the inclusion of EFT truncation uncertainties in the Bayesian inference, the parametric uncertainty is closely related to the actual EFT truncation uncertainty.
For full uncertainty quantification in nuclear structure calculations, the parametric uncertainty must be augmented with additional EFT truncation and many-body method uncertainties.
A key challenge here is that these three uncertainties (parametric, EFT truncation, and many-body method) are all correlated, but this correlation is challenging to study quantitatively and needs further work.
A systematic study of correlations between truncation uncertainties of observables across nuclei remains an important subject for future research.

\section*{Data availability}
The data that support the findings of this
article are openly available~\cite{plies_2026_21130503}.

\begin{acknowledgments}

We thank Christian Forssén, Dick Furnstahl, Hannah Göttling, Takayuki Miyagi, Daniel Phillips, and Isak Svensson for helpful discussions.
This work was supported in part by the European Research Council (ERC) under the European Union's Horizon 2020 research and innovation programme (Grant Agreement No.~101020842),
by the Deutsche Forschungsgemeinschaft (DFG, German Research Foundation) -- Project-ID 279384907 -- SFB 1245, 
by the Laboratory Directed Research and Development Program of Oak Ridge National Laboratory, managed by UT-Battelle, LLC, for the U.S.\ Department of Energy, by the U.S.\ Department of Energy, Office of Science, Office of Nuclear Physics under Award No.~DE-SC0026198,
and by the U.S.\ Department of Energy, Office of Science, Office of Advanced Scientific Computing Research and Office of Nuclear Physics, Scientific Discovery through Advanced Computing (SciDAC) program (SciDAC-5 NUCLEI).
This research used resources of the Oak Ridge Leadership Computing Facility located at Oak Ridge National Laboratory, which is supported by the Office of Science of the Department of Energy under contract No.~DE-AC05-00OR22725.
The authors gratefully acknowledge the Gauss Centre for Supercomputing e.V.\ (www.gauss-centre.eu) for funding this project by providing computing time through the John von Neumann Institute for Computing (NIC) on the GCS Supercomputer JUWELS at Jülich Supercomputing Centre (JSC) and the computing time provided to them on the high-performance computer Lichtenberg II at TU Darmstadt, funded by the German Federal Ministry of Education and Research (BMBF) and the State of Hesse.

\end{acknowledgments}

\bibliography{ref}

\end{document}